\documentclass[reprint,pra,aps]{revtex4-2}

\usepackage[utf8]{inputenc}
\usepackage{graphicx}
\usepackage{verbatim}
\usepackage{dsfont}
\usepackage{epstopdf}
\usepackage{natbib}
\usepackage{xcolor}

\usepackage{amsmath,amsfonts,amssymb,amstext}

\def\bra#1{\mathinner{\langle{#1}|}}
\def\ket#1{\mathinner{|{#1}\rangle}}
\newcommand{\braket}[2]{\langle#1|#2\rangle}

\begin{document}

\title[]{Search and state transfer between hubs by quantum walks}

\author{S. Skoupý and M. \v{S}tefa\v{n}\'{a}k}
\affiliation{Department of Physics, Faculty of Nuclear Sciences and Physical Engineering, Czech Technical University in Prague, B\v rehov\'a 7, 115 19 Praha 1 - Star\'e M\v{e}sto, Czech Republic}

\date{\today}

\begin{abstract}
Search and state transfer between hubs, i.e. fully connected vertices, on otherwise arbitrary connected graph is investigated. Motivated by a recent result of Razzoli et al. (J. Phys. A: Math. Theor. {\bf 55},
265303 (2022)) on universality of hubs in continuous-time quantum walks and spatial search, we extend the investigation to state transfer and also to the discrete-time case. We show that the continuous-time quantum walk allows for perfect state transfer between multiple hubs if the numbers of senders and receivers are close. Turning to the discrete-time case, we show that the search for hubs is successful provided that the initial state is locally modified to account for a degree of each individual vertex. Concerning state transfer using discrete-time quantum walk, it is shown that between a single sender and a single receiver one can transfer two orthogonal states in the same run-time. Hence, it is possible to transfer an arbitrary quantum  state of a qubit between two hubs. In addition, if the sender and the receiver know each other location, another linearly independent state can be transferred, allowing for exchange of a qutrit state. Finally, we consider the case of transfer between multiple senders and receivers. In this case we cannot transfer specific quantum states. Nevertheless, quantum walker can be transferred with high probability in two regimes - either when there is a similar number of senders and receivers, which is the same as for the continuous-time quantum walk, or when the number of receivers is considerably larger than the number of senders. Our investigation is based on dimensional reduction utilizing the invariant subspaces of the respective evolutions and the fact that for the appropriate choice of the loop weights the problem can be reduced to the complete graph with loops.  
\end{abstract}

\maketitle

\section{Introduction}

One of the most promising applications of quantum walks \cite{Aharonov1993,Meyer1996,Farhi1998} is the spatial search
\cite{aaronson_quantum_2003} which can be seen as an extension of the Grover's algorithm \cite{Grover1997} for search of an unsorted database on a quantum computer. Quantum walk search was shown to provide quadratic speed-up over classical search on various graphs and lattices \cite{Shenvi2003,Childs2004,childs_2004b,Ambainis2005,Potocek2009,hein2009:search,reitzner2009,hein_search_lattice_2010,wong2015,wong2018,rhodes2019,rhodes2020,chiang2020,hoyer2020}. In fact, search utilizing continuous-time quantum walk for a single marked vertex was shown to be optimal for almost all graphs \cite{chakraborty2016}. More recently, quadratic speed-up over a classical random walk search for any number of marked vertices was achieved in both discrete-time \cite{ambainis_quadratic_2020} and continuous-time \cite{apers_quadratic_2022} quantum walk algorithms. 

Closely related problem to spatial search is state transfer \cite{bose2003} which is the basic tasks of any quantum communication network \cite{Gisin_QCom_2007}. Apart from direct exchange of quantum information between nodes of a quantum network it can be utilized for distributed quantum computing \cite{Buhrman_DQC_2003,Cuomo_distributed_QC_2020} or more generally in quantum internet \cite{Kimble_QI_2008,Caleffi_QI_communication_2018,Rohde_quantum_internet_2021}. While in search we evolve the system from the equal weight superposition to a localized state on the marked vertex, in state transfer we aim to evolve from a localized state from the sender to the receiver vertex. One possibility to achieve state transfer is to design the dynamics on the whole graph accordingly. For the continuous-time evolution, where the dynamics is governed by the Schroedinger equation with a given Hamiltonian, the problem was over the years investigated on various types of graphs \cite{christandl_perfect_2004,christandl_perfect_2005,plenio_high_2005,bose_2007, kostak_perfect_2007,gualdi_perfect_2008,kay_perfect_2010,kendon2011,godsil_state_2012,nikolopoulos_analysis_2012,hoskovec_decoupling_2014,frydrych_selective_2015,coutinho_perfect_2015,li_qubit_chain_2018,godsil_state_2020,chen_pair_2020,cao_perfect_2021,hoskovec_dynamical_2022,cao_perfect_2022,arezoomand_perfect_2022,wang_perfect_2022,wang_perfect_2023,wang_abstract_2023,arezoomand_perfect_2023,wang_abstract_2023}. Discrete-time version was also studied in detail \cite{kurzynski2011,yalcinkaya2015,shang2018,chen2019,zhan_quantum_2021,kubota_perfect_2022,guo:2022,chan:2023}. Second possibility to achieve state transfer is to employ spatial search, thus altering the dynamics only locally at the sender and the receiver vertex \cite{hein2009}. This approach was extensively investigated especially in the discrete-time quantum walk model \cite{barr2014,stefanak2016,stefanak2017,zhan2019,cao2019,skoupy:2022,Santos_2022,stefanak_hypercube_2023,li_high-fidelity_2023,huang_perfect_2024}.

In a recent paper \cite{Razzoli_hub_ctqw_2022} the authors have investigated continuous-time quantum walk on an arbitrary graph with fully connected vertices, which we refer to as hubs for short. The Hamiltonian of the studied walk was given by the graph Laplacian, with additional potential on marked hubs. It was shown that such quantum walks have certain universal behaviour when the probability amplitudes at the hubs are concerned, guaranteed by the existence of invariant subspaces \cite{novo2015} which are independent of the rest of the graph. Applications to spatial search and quantum transport for single and multiple hubs were discussed, showing that the walk behaves the same as if the graph would be complete.

Our paper elaborates on the ideas of \cite{Razzoli_hub_ctqw_2022} and extends the investigation to state transfer between multiple hubs and also to the discrete-time scenario. Utilizing the effective Hamiltonian of the continuous-time quantum walk with marked hubs derived in \cite{Razzoli_hub_ctqw_2022}, we show that state transfer from $S$ senders to $R$ receivers can be achieved with high fidelity provided that $S\approx R$. Moving on to the discrete-time case, we first investigate a spatial search for $M$ marked hubs. The dynamics is given by a coined walk with a flip-flop shift. As the quantum coin we consider a modified Grover coin with a weighted loop at each vertex \cite{wong2015,wong2018,rhodes2019,rhodes2020,chiang2020,hoyer2020}. The hubs which are a solution to the search problem are marked by an additional phase shift of $\pi$. We show that there is a five-dimensional invariant subspace of the walk provided that the initial state of the search is locally modified such that it has equal projection onto every vertex subspace. In this case the walk evolves as search on a complete graph with loops, which is known to be optimal. We provide a detailed investigation how do the terms contributing to the total success probability change with the number of solutions $M$. Next, we proceed to the state transfer between two marked hubs. Properly choosing the weights of the loops of the local coins turns the problem of state transfer between two hubs on an otherwise arbitrary graph to the state transfer on a complete graph with loops, which we have investigated earlier \cite{stefanak2016}. We expand the known results and show that it is possible to transfer multiple orthogonal states. We find that there is a 9-dimensional invariant subspace, which can be further split into symmetric and antisymmetric subspaces with respect to the exchange of the sender and the receiver vertices. Time evolution in the invariant subspace is investigated in detail for two orthogonal initial states, namely loop and the equal weight superposition of all outgoing arcs at the sender vertex. It is shown that in both cases the discrete-time quantum walk achieves transfer to the corresponding state at the receiver vertex in the same run-time. Hence, we can transfer an arbitrary quantum state of a qubit between the hubs. Since our analysis utilizes an approximation of effective evolution operator eigenstates for large graphs, we also numerically investigate the fidelity of qubit state transfer in dependence on the number of vertices $N$. It is shown that the fidelity behaves as $1-O(1/N)$. In addition, we show that if the sender and the receiver know each other's location, a third linearly independent state can be transferred in the same run-time, namely the arc on the edge connecting the sender and the receiver. Hence, in this case the sender and the receiver can exchange a qutrit state. Finally, we extend the study to state transfer from $S$ sender hubs to $R$ receiver hubs. For $S\neq R$ the exchange symmetry between senders and receivers is broken and the splitting of the invariant subspace (which is now 11-dimensional) into symmetric and anti-symmetric subspaces is no longer possible. Nevertheless, the system can be investigated analytically. We show that while we can't send a qubit state, transfer of the walker from senders to receivers occurs with high probability in two regimes - either when $S\approx R$ as for the continuous time case, or when the number of receivers is considerably larger than the number of senders.

The rest of the paper is organized as follows: in Section~\ref{sec:ctqw:sta} we review the results of \cite{Razzoli_hub_ctqw_2022} for the continuous-time quantum walk and extend the investigation to state transfer between multiple hubs. Section~\ref{sec:sa} introduces the discrete-time quantum walk and investigates search for $M$ marked hubs. Section~\ref{sec:sta} is dedicated to the state transfer between two hubs by discrete-time quantum walk. In Section~\ref{sec:sta:multi} we investigate state transfer between multiple hubs. We conclude and present an outlook in Section~\ref{sec:concl}. Technical details concerning the form of the individual evolution operators in the invariant subspaces and their eigenstates are left for Appendices.

\section{State transfer between hubs by continuous-time quantum walk}
\label{sec:ctqw:sta}

We begin with the continuous-time quantum walk. Let us have a graph $G=(V,E)$, where $V$ denotes the set of vertices and $E$ denotes the set of edges. We consider simple graphs, i.e. the are no loops and no multiple edges in $E$. For a graphs with $N=|V|$ vertices the Hilbert space of the continuous-time walk is an $N$-dimensional complex space spanned by vectors $\ket{v}$, $v\in V$, corresponding to a particle being localized at the vertex $v$. The authors of \cite{Razzoli_hub_ctqw_2022} consider $M$ vertices $w\in W$ to be marked hubs, i.e. fully connected with degrees $d_w = N-1$. The Hamiltonian of the walk is taken as the weighted Laplacian with additional potential on the marked vertices
\begin{equation}
    \hat H = \gamma \hat L + \sum_{w\in W}\lambda_w \ket{w}\bra{w}.
\end{equation}
Note that the Laplacian of a given graph $G$ is determined by its adjacency matrix $\hat A$ and the diagonal degree matrix $\hat D$ as
\begin{equation}
    \hat L = \hat D - \hat A.
\end{equation}
The results of \cite{Razzoli_hub_ctqw_2022} show that as far as the amplitudes at the marked vertices are concerned, the system evolves in a $M+1$ invariant subspace \cite{novo2015} spanned by $\ket{w}$, $w\in W$, and the equal weight superposition of all remaining vertices 
\begin{equation}
\label{ctqw:g}
   \ket{g} = \frac{1}{\sqrt{N-M}}\sum_{v\notin W} \ket{v} . 
\end{equation}
The effective Hamiltonian in the invariant subspace acts according to (see Eq. (50) in \cite{Razzoli_hub_ctqw_2022} for $\mu=M$)
\begin{eqnarray}
\nonumber \hat H\ket{w} & = & \gamma\left(N-1+\frac{\lambda_w}{\gamma}\right)\ket{w}  - \gamma\sum_{\substack{w \in W\\w'\neq w}} \ket{w'} - \\
\nonumber & & - \gamma\sqrt{N-M} \ket{g}, \quad \forall w\in W,\\
\hat H \ket{g} & = & -\gamma\sqrt{N-M}\sum_{w\in W} \ket{w} + \gamma M \ket{g}. 
\label{ctqw:h}
\end{eqnarray}
For spatial search, the weights of all marked hubs are chosen as $\lambda_w=-1$, and the optimal choice of the hopping amplitude $\gamma$ is shown to be equal to $\frac{1}{N}$. The dimensionality of the problem can be further reduced to 2, since the target state of the search algorithm is the equal weight superposition of the marked vertices
\begin{equation}
    \ket{\tilde{w}} = \frac{1}{\sqrt{M}}\sum_{w\in W}\ket{w}.
\end{equation}
In the invariant subspace spanned by $\ket{\tilde{w}}$ and $\ket{g}$ the Hamiltonian is described by a matrix (see Eq. (72) in \cite{Razzoli_hub_ctqw_2022} for $\gamma = \frac{1}{N}$)
\begin{equation}
    H = -\frac{1}{N} \begin{pmatrix}
        M & \sqrt{M(N-M)} \\
        \sqrt{M(N-M)} & -M 
    \end{pmatrix}.
\end{equation}
Search evolves from equal weight superposition of all vertices 
\begin{equation}
\label{ctqw:init:search}
    \ket{\psi} = \frac{1}{\sqrt{N}}\sum_{v}\ket{v} = \sqrt{\frac{M}{N}}\ket{\tilde{w}} + \sqrt{\frac{N-M}{N}}\ket{g},
\end{equation}
towards the target state $\ket{\tilde{w}}$ with unit probability in time given by
\begin{equation}
    T = \frac{\pi}{2}\sqrt{\frac{N}{M}}.
    \label{time:search:ctqw}
\end{equation}

Let us now utilize the Hamiltonian (\ref{ctqw:h}) for state transfer. Consider $S$ sender and $R$ receiver hubs from sets $\cal S$ and $\cal R$, such that $S+R=M$. We keep the hopping rate $\gamma = \frac{1}{N}$ and the weights $\lambda_s = \lambda_r = -1$ for all $s\in {\cal S}$ and $r\in {\cal R}$. In such a case, we can group together the states localized on the sender and the receiver vertices
\begin{eqnarray}
    \nonumber \ket{\cal S} & = & \frac{1}{\sqrt{S}} \sum_{s\in{\cal S}} \ket{s}, \\
    \ket{\cal R} & = & \frac{1}{\sqrt{R}} \sum_{r\in{\cal R}} \ket{r},
\end{eqnarray}
which we consider as the initial and the target state of state transfer. Together with (\ref{ctqw:g}) they form a 3D invariant subspace. 
The Hamiltonian in the reduced subspace is given by the following 3x3 matrix
\begin{equation}
H = -\frac{1}{N} \begin{pmatrix}
    S & \sqrt{R S} & \sqrt{S(N-M)}\\
    \sqrt{R S} & R  & \sqrt{R(N-M)} \\
   \sqrt{S(N-M)} & \sqrt{R(N-M)} & -M 
    \end{pmatrix}    
\label{hamiltonian}
\end{equation}
The energy spectrum of (\ref{hamiltonian}) is found to be
\begin{equation}
\label{ctqw:energy:r=1}
E_0 = 0,\quad  E_\pm = \pm \sqrt{\frac{M}{N}},    
\end{equation}
and the corresponding eigenvectors are
\begin{eqnarray}
\label{ctqw:evec:r=1}
\ket{0} & = & \frac{1}{\sqrt{S+R}}(\sqrt{R}\ket{\cal S} - \sqrt{S}\ket{\cal R}), \\
\nonumber \ket{\pm} & = & \sqrt{\frac{1- E_\pm}{2M}} \left(\sqrt{S}\ket{\cal S} + \sqrt{R}\ket{\cal R} \right) \mp \\
\nonumber & & \mp \sqrt{\frac{1+ E_\pm}{2}}\ket{g}.
\end{eqnarray}

\begin{figure}
    \centering
    \includegraphics[width=0.45\textwidth]{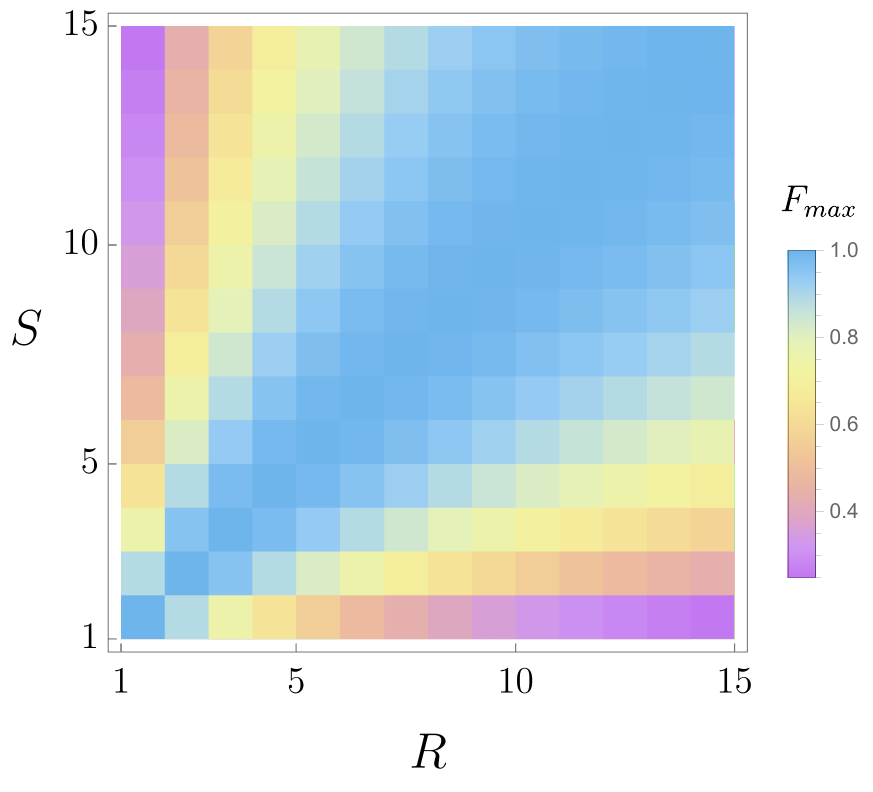}
    \caption{Maximal state transfer fidelity (\ref{ctqw:fmax}) in continuous time quantum walk between $S$ senders and $R$ receivers.}
    \label{fig:ctqw:transfer:max:sr}
\end{figure}

Given the initial state $\ket{\psi(0)} = \ket{\cal S}$, the state of the walk at time $t$ reads
\begin{eqnarray}
\nonumber \ket{\psi(t)} & = & e^{-i H t}\ket{\cal S} \\
\nonumber & = & \sqrt{\frac{R}{M}}\ket{0} +   e^{-i E_+ t}\sqrt{\frac{S(1-E_+)}{2 M}} \ket{+} + \\
& & + e^{i E_+ t}\sqrt{\frac{S(1+E_+)}{2M}} \ket{-}.
\end{eqnarray}
Overlap with the target state $\ket{\cal R}$ then equals
\begin{equation}
\braket{{\cal R}}{\psi(t)} = \frac{\sqrt{RS}}{M}\left[1-\cos{(E_+ t)} + i E_+ \sin{(E_+ t)} \right].
\end{equation}
Fidelity of state transfer is then given by
\begin{eqnarray}
\nonumber F(t) & = & |\braket{{\cal R}}{\psi(t)}|^2 \\
\label{fid:ctqw}   & = & \frac{4 RS}{M^2}\sin^4{\left(\frac{E_+ t}{2} \right)} + \frac{R S}{N M} \sin^2{(E_+ t)}.
\end{eqnarray}
Hence, the maximal fidelity 
\begin{equation}
\label{ctqw:fmax}
    F_{max} = \frac{4RS}{(R+S)^2},
\end{equation}
is reached at time $T$ given by
\begin{equation}
    T = \frac{\pi}{E_+} = \sqrt{\frac{N}{R+S}}\pi.
\end{equation}
We find that state transfer with high fidelity is possible when $R\approx S$, see Figure~\ref{fig:ctqw:transfer:max:sr} for a visualization. Unit fidelity corresponding to perfect state transfer is achieved when the number of receivers and senders are equal. Run-time of the state transfer is twice of that for search for $M=R+S$ vertices (\ref{time:search:ctqw}). Indeed, in state transfer we evolve the system from the state localized on the sender vertices through the equal weight superposition of all vertices (\ref{ctqw:init:search}) towards the state localized on the receiver vertices, effectively iterating the search dynamics twice. For $S$ and $R$ differing considerably the state transfer is suppressed. In particular, for $S\gg R$ or $R\gg S$, maximal fidelity (\ref{ctqw:fmax}) tends to zero. 

\section{Discrete-time quantum walk search for hubs}
\label{sec:sa}

We turn to the discrete-time quantum walk search for $M$ marked hubs labeled by indices from a set $\cal M$. Let us start with the Hilbert space. Given a graph $G=\left(V,E\right)$ the corresponding Hilbert space ${\cal H}_G$ can be decomposed as a direct sum of local Hilbert spaces ${\cal H}_v$ at each vertex $v\in V$. Each ${\cal H}_v$ is spanned by vectors $\ket{v,w}$ such that there is an edge between vertex $v$ and $w$, and an additional state $\ket{v,v}$ corresponding to the loop at the vertex $v$. The dimension of the local space ${\cal H}_v$ is thus equal to $d_v + 1$, where $d_v$ is the degree of the vertex $v$ in the simple graph $G$.

The unitary evolution operator of the walk, $\hat U = \hat S \hat C$, is a product of the shift operator $\hat{S}$ and the coin operator $\hat{C}$. We consider the flip-flop shift, which is defined in the following way 
\begin{eqnarray}
\label{shift:search}
\hat{S}\vert v,w\rangle = \vert w,v\rangle, \quad \hat S\ket{v,v} = \ket{v,v} .
\label{shift}
\end{eqnarray}
The coin operator is given by a direct sum
\begin{equation}
\hat C = \bigoplus_v \hat C_v,    
\end{equation}
of local unitaries $\hat C_v$ acting on vertex spaces ${\cal H}_v$. As the local coins at non-marked vertices we consider the Grover operator with a weighted loop \cite{wong2018}. The original Grover operator \cite{Grover1997} is invariant under all permutations. Hence, when used as a coin in the discrete time quantum walk, it has the advantage that it does not matter how we order the basis states in the local Hilbert space ${\cal H}_v$ \cite{Shenvi2003}. Adding the weighted loop adds a parameter, which can be tuned to improve the success probability of search \cite{hoyer2020}. We can write the Grover operator with a weighted loop as \cite{skoupy:2022}
\begin{eqnarray}
\hat{G}_v(l_v)= 2\vert\Omega_v(l_v)\rangle\langle\Omega_v(l_v)\vert -\hat{I}_v ,
\label{grover}
\end{eqnarray}
where $\hat{I}_v$ is an identity operator at the subspace ${\cal H}_v$ and $\vert\Omega_v(l_v)\rangle$ is given by
\begin{equation}
\ket{\Omega_{v}(l_v)}=\frac{1}{\sqrt{d_v+l_v}}\left(\sqrt{d_v}\ket{\Omega_v} + \sqrt{l_v} \ket{v,v}\right) .    
\end{equation}
By $\ket{\Omega_v}$ we have denoted the equal weight superposition of all outgoing arcs from the vertex $v$
\begin{eqnarray}
\vert\Omega_v\rangle=\frac{1}{\sqrt{d_v}}\sum_{\substack{w\\ \left\lbrace v,w\right\rbrace\in E}}\vert v,w\rangle
\label{localSuperposition}.
\end{eqnarray}
The free parameter $l_v$ sets the weight of the loop state at the vertex $v$. In correspondence to the continuous-time case \cite{Razzoli_hub_ctqw_2022} we tune it according to the degree of the vertex and set $l_v=N-d_v$. On the marked hubs $m\in{\cal M}$ this results in $l_m = 1$. In addition, we multiply the coin on the marked hubs by -1, which is analogous to the marking of the solutions in Grover algorithm \cite{Grover1997}. Moreover, one can show that the phase shift of $\pi$ is the optimal choice, as it results in large gap of the avoided crossing which is inversely proportional to the run-time of the search \cite{hein2009:search,hein_search_lattice_2010}. Hence, choosing different phase shift results in slower algorithm with lower success probability. The evolution operator of the search is then given by
\begin{equation}
\label{dtqw:evol:search}
    \hat U_{\cal M} = \hat S\ \left(\bigoplus_{v\notin {\cal M}} \left(\hat G_v(N-d_v)\right)\bigoplus_{m\in {\cal M}}  (-\hat{G}_m(1))\right) .
\end{equation}

The usual choice of the initial state for the search algorithm is the equal weight superposition of all arcs
\begin{equation}
\vert \tilde{\Omega}\rangle = \frac{1}{\sqrt{\sum\limits_{v\in V}
d_v}}\sum_{v\in V} \sqrt{d_v}\vert\Omega_v\rangle .
\label{sa:init:old}
\end{equation}
However, with this initial state the search does not perform well, as can be observed from numerical simulations. In addition, we do not find a simple closed invariant subspace of $\hat U_{\cal M}$ which involves (\ref{sa:init:old}). The reason for that is that $\ket{\tilde{\Omega}}$ has the same overlap with every arc, but not with every vertex of the graph, since the graph is not expected to be regular. Therefore, state (\ref{sa:init:old}) is not an eigenstate of the unperturbed walk operator (i.e. without marking of hubs by a $\pi$ phase shift), which is given by 
\begin{equation}
\hat U = \hat S \left(\bigoplus_{v\in V} \left(\hat G_v(N-d_v)\right) \right).  
\end{equation}
Such eigenstate reads
\begin{eqnarray}
\ket{\Omega} = \frac{1}{\sqrt{N}}\sum_{v\in V}\ket{\Omega_v(N-d_v)},
\label{newSearchInit}
\end{eqnarray}
which has the same overlap of $\frac{1}{\sqrt{N}}$ with all vertex subspaces $\mathcal{H}_v$. One can easily check that it satisfies $\hat U \ket{\Omega} = \ket{\Omega}$. For this initial state we find that search for hubs behaves analogously as search on the complete graph with loops \cite{Ambainis2005,wong2015}. Let us construct the invariant subspace \cite{reitzner2009,stefanak2016} of the evolution operator (\ref{dtqw:evol:search}). First, we include states where the walker is localized on the marked hubs 
\begin{eqnarray}
        \label{basis:search1} \ket{\nu_1} &=& \frac{1}{\sqrt{M}}\sum\limits_{m\in{\cal M}}\ket{m,m} , \\
	\nonumber \ket{\nu_2} &=& \frac{1}{\sqrt{M(M-1)}}\sum\limits_{m\neq m'\in{\cal M}}\ket{m,m'},\\
	\nonumber \ket{\nu_3} &=& \frac{1}{\sqrt{M(N-M)}}\sum\limits_{m\in{\cal M}}\sum\limits_{v\notin{\cal M}}\ket{m,v},
\end{eqnarray}
corresponding to superposition of loops ($\ket{\nu_1}$), arcs of the edges connecting different marked hubs ($\ket{\nu_2}$), and arcs leaving marked hubs to the rest of the graph ($\ket{\nu_3}$). Since the evolution operator (\ref{dtqw:evol:search}) treats all marked hubs equally we can group them together and form the superpositions of the corresponding internal states. Note that the flip-flop shift operator (\ref{shift:search}) leaves the states $\ket{\nu_1}$ and $\ket{\nu_2}$ unchanged, however, it maps the vector $\ket{\nu_3}$ onto
\begin{eqnarray}
\nonumber    \label{basis:search2} \ket{\nu_4} &=& \hat S\ket{\nu_3} =  \frac{1}{\sqrt{M(N-M)}}\sum\limits_{m\in{\cal M}}\sum\limits_{v\notin{\cal M}}\ket{v,m}, \\
\end{eqnarray}
where the walker is localized on the non-marked vertices in the arcs heading towards the marked ones. We add this state as the fourth vector of the invariant subspace $\cal I$. To complete the invariant subspace we consider the superposition of loops and arcs between non-marked vertices
\begin{eqnarray}
\label{basis:search3}
\nonumber \ket{\nu_5} &=& \frac{1}{N-M}\sum\limits_{v\notin{\cal M}}\left(\sqrt{N}\ket{\Omega_v(N-d_v)}-\sum\limits_{m\in{\cal M}}\ket{v,m}\right). \\
\end{eqnarray}
These 5 vectors are required to decompose the initial state of the search (\ref{newSearchInit}) according to
\begin{eqnarray}
\label{init:search:2}
    \ket{\Omega} & = & \frac{1}{N}\left(\sqrt{M}\ket{\nu_1}+\sqrt{M(M-1)}\ket{\nu_2}  + \right. \\
    \nonumber & & + \left. \sqrt{M(N-M)}(\ket{\nu_3} + \ket{\nu_4}) + (N-M)\ket{\nu_5}\right).
\end{eqnarray}
We note that in search for a single marked hub $(M=1)$ the state $\ket{\nu_2}$ vanishes and the invariant subspace $\cal I$ is only 4-dimensional.

The action of the search evolution operator (\ref{dtqw:evol:search}) on the states $\ket{\nu_j}$ is described in the   Appendix \ref{app:a}, see Eq. (\ref{evol:op:search}). We find that the spectrum of $\hat U_{\cal M}$ reduced to the invariant subspace is composed of $\pm 1$ and pair conjugate eigenvalues $\lambda_{\pm}=e^{\pm i\omega}$ where the phase is given by
\begin{eqnarray}
\omega = \arccos\left(1-\frac{2M}{N}\right).
\label{eigenphase2}
\end{eqnarray}
For $M>1$ eigenvalue 1 is doubly degenerate. We denote the corresponding eigenvectors as $\ket{(1)_{1,2}}$, $\ket{-1}$ and $\ket{\pm\omega}$, their explicit forms are given in the formulas (\ref{search:ev:p1}), (\ref{search:ev:m1}) and (\ref{search:ev:omega}.)
Decomposition of the initial state (\ref{init:search:2}) into the eigenbasis of $\hat U_{\cal M}$ is given by
\begin{equation}
\ket{\Omega}  =   - \sqrt{\frac{N-M}{2N}} \ket{(1)_1} + \sqrt{\frac{M}{2N}}\ket{-1} + \frac{1}{2}(\ket{+\omega} + \ket{-\omega}). 
\end{equation}
Evolution of the walk after $t$ steps then can be written as 
\begin{eqnarray}
\nonumber \ket{\psi(t)} & = & - \sqrt{\frac{N-M}{2N}}\ket{(1)_1} + (-1)^t\sqrt{\frac{M}{2N}} \ket{-1} + \\
& & + \frac{1}{2}(e^{i\omega t}\ket{+\omega} + e^{-i\omega t}\ket{-\omega}).
\end{eqnarray}

The total probability to find one of the marked hubs after $t$ steps can be written as a sum of transition probabilities from $\ket{\psi(t)}$ into the basis states $\ket{\nu_i}$, $i = 1,2,3$
\begin{equation}
    P(t) = \sum_{i=1}^3 P_i(t) = \sum_{i=1}^3 |\braket{\nu_i}{\psi(t)}|^2 .
\end{equation}
We find that the corresponding probability amplitudes read
\begin{eqnarray}
\nonumber \braket{\nu_1}{\psi(t)} & = & \frac{1}{2\sqrt{M}}\left(\cos{(\omega t)} - 1 + (1+(-1)^t)\frac{M}{N}\right), \\
\braket{\nu_2}{\psi(t)} & = & \sqrt{M-1}\braket{\nu_1}{\psi(t)} , \\
\nonumber \braket{\nu_3}{\psi(t)} & = & \frac{1}{2}\sin{(\omega t)}  + (1+(-1)^t)\frac{\sqrt{M(N-M)}}{2N} .
\end{eqnarray}
Considering first the case $N\gg M$, we can neglect the rapidly oscillating terms and obtain the transition probabilities
\begin{eqnarray}
\nonumber    P_1(t) & \approx & \frac{1}{M}  \sin^4\left(\frac{\omega t}{2}\right), \\
\nonumber P_2(t) & \approx &  \frac{M-1}{M}\sin^4\left(\frac{\omega t}{2}\right) , \\
P_3(t) & \approx & \frac{1}{4}\sin^2(\omega t) . 
\label{trans:i:small:M}
\end{eqnarray}
The total success probability is then equal to
\begin{equation}
\label{search:succ:small:M}
    P(t) \approx \sin^4\left(\frac{\omega t}{2}\right) + \frac{1}{4}\sin^2(\omega t),
\end{equation}
which is close to unity provided that we choose the number of steps $t$ as the closest integer to
\begin{eqnarray}
T = \frac{\pi}{\omega} \approx\frac{\pi}{2}\sqrt{\frac{N}{M}}+O\left(\sqrt{\frac{M}{N}}\right) .
\label{stepNumber2}
\end{eqnarray}
For $N\gg M$ the maximum success probability comes from $P_1(T)$ and $P_2(T)$, while $P_3(T)$ vanishes. When $M=1$ we find the particle in the loop at the unique marked vertex, since there is no $\ket{\nu_2}$ state and $P_2(t)$ equals zero. As the number of marked hubs grows the contribution of $P_2(t)$ increases, so we are more likely to find the walker on an edge connecting two marked hubs.

When the number of marked hubs is non-negligible relative to the total number of vertices the approximations (\ref{trans:i:small:M}) are no longer valid. While $P_1(t)$ vanishes the contribution of $P_3(t)$ to the total success probability becomes more important. One can show that for all values of $N$ and $M<N$ the total success probability of search evolves with the number of steps according to
\begin{equation}
    P(2t) = P(2t+1) = \sin^2{\left(\frac{\omega (2t+1)}{2}\right)}.
    \label{succ:prob:sa}
\end{equation}
Indeed, as was shown in \cite{Ambainis2005} two steps of the quantum walk search on a complete graph with loops are equivalent to a single iteration of the Grover algorithm. In conclusion, we find one of the marked hubs with probability close to one for the number of steps derived earlier (\ref{stepNumber2}).

\begin{figure}
\centering
\includegraphics[width=0.45\textwidth]{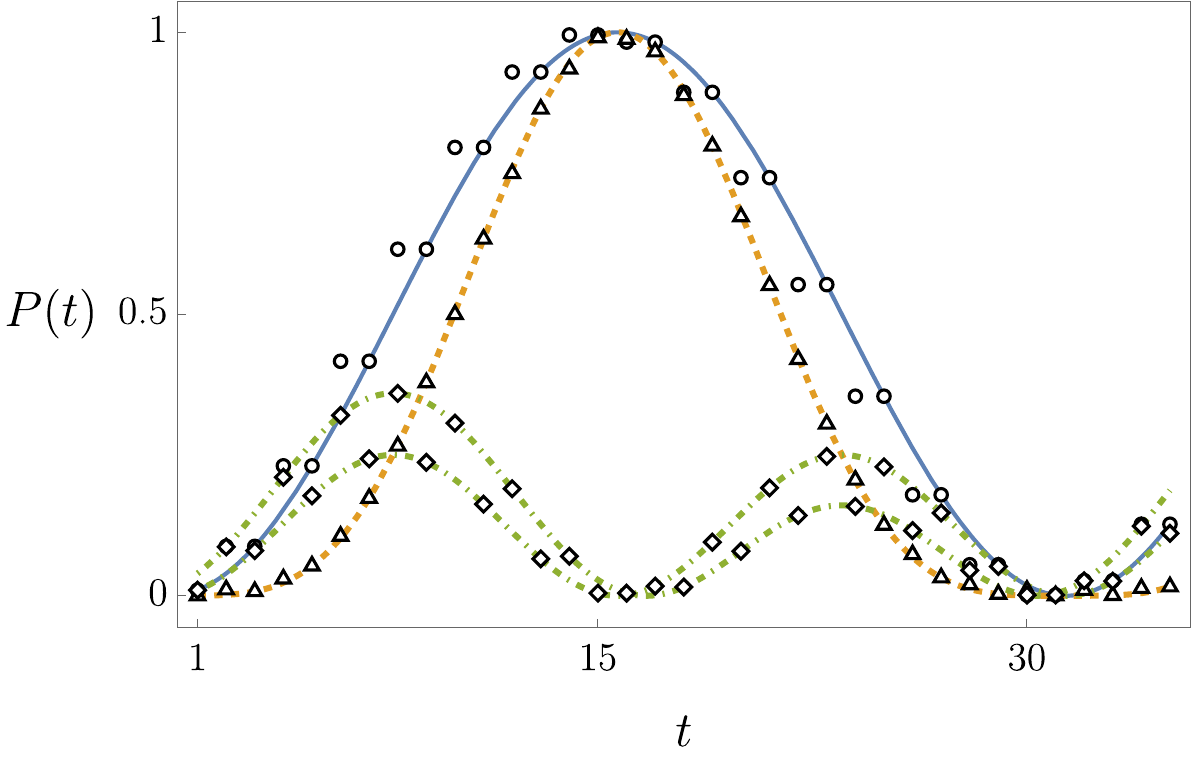}
\includegraphics[width=0.45\textwidth]{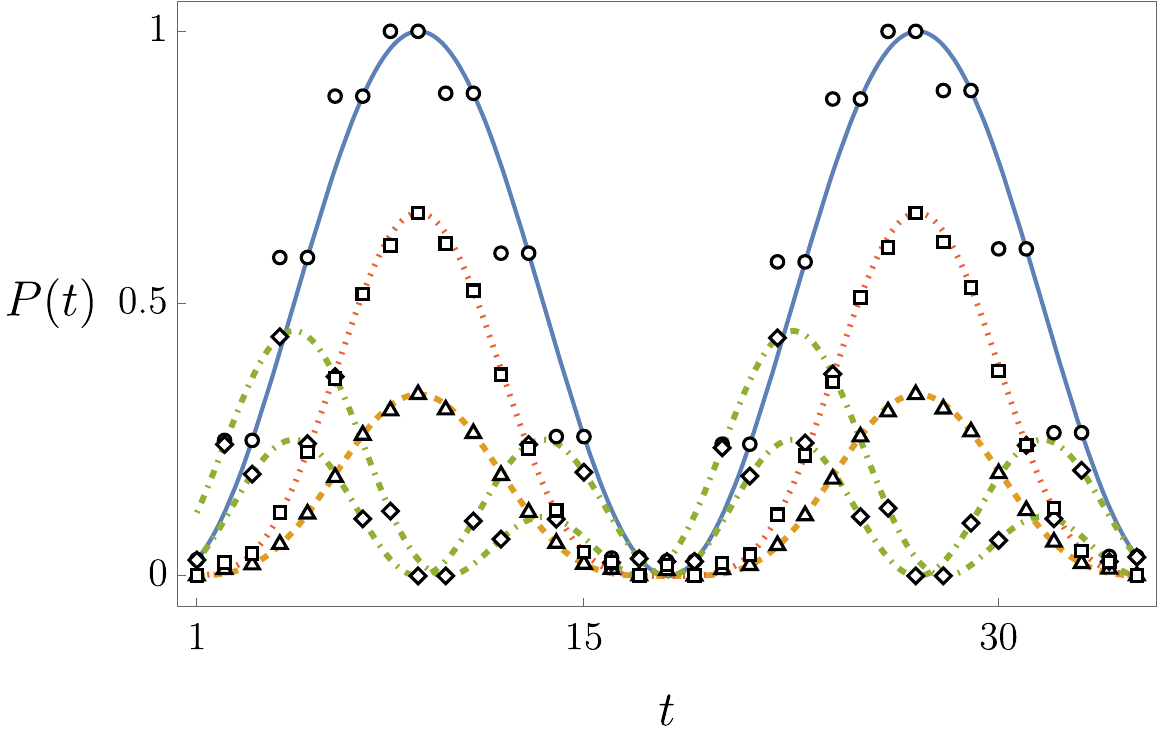}
\includegraphics[width=0.45\textwidth]{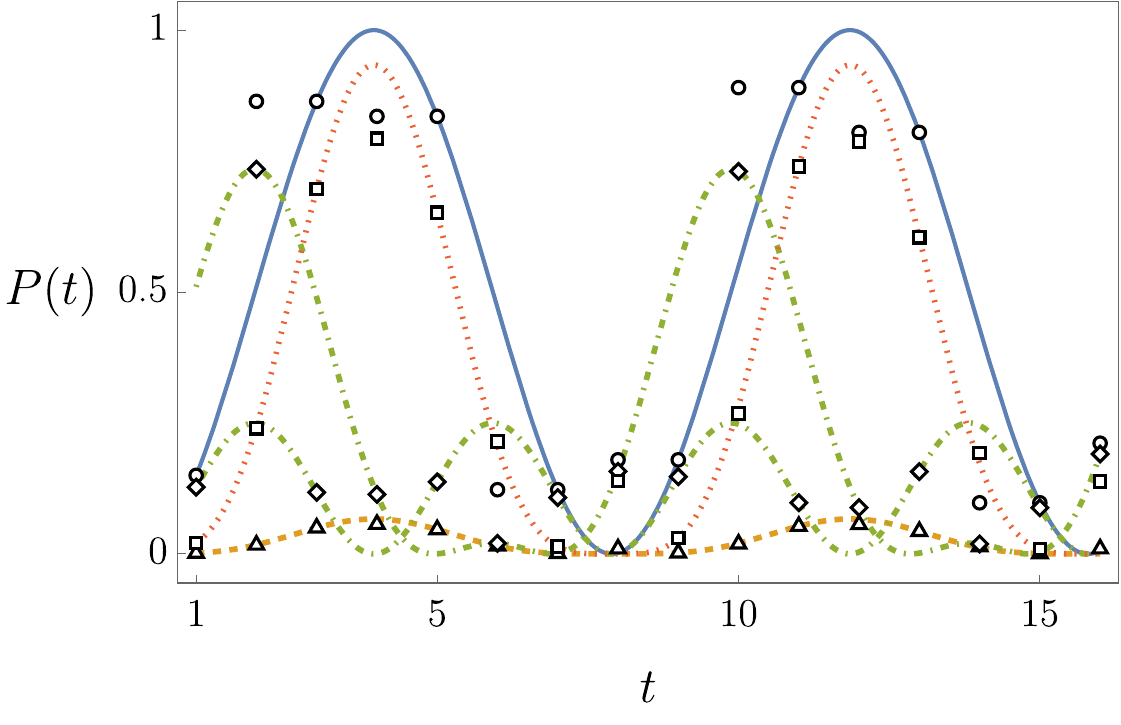}
\caption{Evolution of the success probability of search for a different number of hubs $M$ on a graph with $N=100$ vertices. Data points are from numerical simulation, curves are given by the formulas (\ref{trans:i:small:M}) and (\ref{search:succ:small:M}). In the upper plot we have chosen $M=1$, in the middle plot we consider $M=3$ and the bottom plot shows the results for $M=15$. Triangles and the dashed yellow curve correspond to $P_1(t)$, squares and the orange dotted curve depict $P_2(t)$, and diamonds and the dot-dashed green curves (one for odd and one for even steps) show $P_3(t)$. Circles and the full blue curve depict the total success probability $P(t)$.}
\label{fig:sa}
\end{figure}

For comparison of analytical and numerical results see Fig. \ref{fig:sa}, where we display the course of the success probability on a graph with $N = 100$ vertices with different number of marked hubs. The upper plot shows the case of a single marked hub, i.e. $M=1$. In the middle plot we have considered $M=3$, and the lower plot shows the case $M=15$. The plot highlights how the contribution of different terms $P_i(t)$ changes as the number of marked hubs increases. One can also notice that the total success probability changes only after each two steps, in accordance with (\ref{succ:prob:sa}).

\section{State transfer between two hubs}
\label{sec:sta}

Let us now turn to the state transfer between two hubs, sender and receiver vertices $s$ and $r$. We consider the same local coins for the marked and non-marked vertices as for the search. The evolution operator for the state transfer then has the form
\begin{eqnarray}
\nonumber \hat U_{s,r} & = & \hat S\ \left(\bigoplus_{v\neq s,r} \left(\hat G_v(N-d_v)\right) \oplus  (-\hat{G}_s(1)) \right. \\
& & \hspace{48pt} \left. \oplus  (-\hat{G}_r(1))\right) .
\label{evol:op:2hubs}
\end{eqnarray}
As for the search, the choice of the loop weights allows to map the problem of state transfer between hubs on otherwise arbitrary graph to state transfer on complete graph with loops. We have investigated this model earlier \cite{stefanak2016}, however, only a single initial state was considered. In the present paper we extend this investigation and show that it is possible to transfer two orthogonal states, namely the loop and the equal weight superposition of all outgoing arcs, in the same run-time. Hence, one can transfer a state of a qubit from one hub to another. In fact, we show that if the sender and the receiver knows their locations, then they can exchange another linearly independent state, allowing for exchange of a qutrit.

First, we perform the dimensional reduction, i.e. determine the invariant subspace ${\cal I}$ of the walk evolution operator $\hat{U}_{s,r}$ containing both pairs $\ket{s,s}$, $\ket{r,r}$ and $\ket{\Omega_s}$, $\ket{\Omega_r}$. Construction of the invariant subspace follows the same logic as for the search problem in the previous Section. We begin with the loops on the marked vertices and the connecting arcs
\begin{align}
\label{s:r:inv1} \ket{\nu_1} & = \ket{s,s} ,\qquad \ket{\nu_2}  = \ket{r,r},\\
\nonumber\vert\nu_3\rangle & = \ket{s,r} ,\qquad  \vert\nu_4\rangle = \ket{r,s}.
\end{align}
Next, we add states connecting sender and receiver with the rest of the graph
\begin{align}
\nonumber \vert\nu_5\rangle & = \frac{1}{\sqrt{N-2}}\sum\limits_{v\neq s,r}\ket{s,v} ,\quad  \vert\nu_6\rangle  = \frac{1}{\sqrt{N-2}}\sum\limits_{v\neq s,r}\ket{r,v}, \\
\nonumber \vert\nu_7\rangle  &= \frac{1}{\sqrt{N-2}}\sum\limits_{v\neq s,r}\ket{v,s}, \quad \vert\nu_8\rangle  = \frac{1}{\sqrt{N-2}}\sum\limits_{v\neq s,r}\ket{v,r}. \\
\label{s:r:inv2}
\end{align}
To complete the invariant subspace $\cal I$ we consider vectors corresponding to loops and arcs between non-marked vertices
\begin{align}
\nonumber \vert\nu_9\rangle  & =  \frac{1}{N-2}\sum\limits_{v\neq s,r}\left(\sqrt{N}\ket{\Omega_v(N-d_v)}  -  \ket{v,r}-\ket{v,s}\right) . \\
\label{s:r:inv3}
\end{align}
Note that the equal weight superpositions can be expressed in the form
\begin{eqnarray}
\nonumber \ket{\Omega_s} & = & \frac{1}{\sqrt{N-1}}\ket{\nu_3} + \sqrt{\frac{N-2}{N-1}}\ket{\nu_5}, \\
\label{omega:sr} \ket{\Omega_r} & = & \frac{1}{\sqrt{N-1}}\ket{\nu_4} + \sqrt{\frac{N-2}{N-1}}\ket{\nu_6},
\end{eqnarray}
which tends to $\ket{\nu_5}$ and $\ket{\nu_6}$ for large graph size $N$. One can show by direct calculation that the 9-dimensional space is closed under action of $\hat U_{s,r}$. We note that in \cite{stefanak2016} we considered state transfer on the complete graph with loops from the state corresponding to the equal weight superposition of the loop and arcs leaving the sender vertex, which in the present notation reads
\begin{equation}
\label{sr:init:old}
    \ket{\psi} = \frac{1}{\sqrt{N}} \left(\ket{\nu_1} + \ket{\nu_3} + \sqrt{N-2}\ket{\nu_5}\right) .
\end{equation}
For this particular initial state it was sufficient to consider a smaller 5 dimensional invariant subspace \cite{stefanak2016}.

We provide the detailed investigation of the evolution operator (\ref{evol:op:2hubs}) in the invariant subspace in Appendix \ref{app:b}. We find that the spectrum of the reduced operator consists of $\pm 1$ and three conjugated pairs $\lambda_j^{(\pm)} = e^{\pm i \omega_j}$ where the phases $\omega_j$ are given by
\begin{eqnarray}
\label{sta:2:omega1}    \omega_1 & = & \arccos\left(1-\frac{4}{N}\right), \\
\label{sta:2:omega23} \omega_2 &=& \arccos\left(\sqrt{1-\frac{2}{N}}\right) = \frac{\omega_1}{2}, \\
\nonumber  \omega_3 &=& \arccos\left(-\sqrt{1-\frac{2}{N}}\right) = \pi - \omega_2 .
\end{eqnarray}
Eigenvalue $1$ is doubly degenerate. The corresponding eigenstates are denoted by $\ket{(1)_{1,2}}$, $\ket{-1}$ and $\ket{\pm \omega_j}$, $j=1,2,3$. Their explicit forms are given by equations (\ref{sta:ev:p1}), (\ref{sta:ev:m1}), (\ref{sta:ev:omega1}), (\ref{sta:ev:omega2}) and (\ref{sta:ev:omega3}). We proceed with the calculation of the time evolution of the walk for different choices of the initial and target states.

\subsection{Transfer of equal weight superposition states}
\label{sec:sta:sup}

Let us first consider the initial state as the equal superposition of all outgoing arcs at the sender vertex (\ref{localSuperposition}). As follows from (\ref{omega:sr}), for large graph the initial and the target states, $\ket{\Omega_s}$ and $\ket{\Omega_r}$, can be approximated by
\begin{eqnarray}
\ket{\Omega_s} & \approx & \ket{\nu_5} ,\quad  \ket{\Omega_r}  \approx \ket{\nu_6}  .
\end{eqnarray}
In terms of the eigenbasis of $\hat{U}_{s,r}$ the vectors $\ket{\Omega_s}$ and $\ket{\Omega_r}$ can be estimated by
\begin{eqnarray}
	\nonumber \ket{\Omega_s} &\approx & \frac{1}{2\sqrt{2}}\left(\sqrt{2}\ket{-1} + i\left(\ket{+\omega_1}-\ket{-\omega_1}\right) + \right.\\
\nonumber  & & \hspace{12pt} \left. \frac{}{} +i\left(\ket{+\omega_2} - \ket{-\omega_2}\right)  - i\left(\ket{+\omega_3} - \ket{-\omega_3}\right)\right),\\
  \ket{\Omega_r} &\approx& \frac{1}{2\sqrt{2}}\left(\sqrt{2}\ket{-1} + i\left(\ket{+\omega_1}-\ket{-\omega_1}\right) - \right. \\
  \nonumber   & & \hspace{12pt} \left. \frac{}{} - i \left(\ket{+\omega_2} - \ket{-\omega_2}\right) + i \left(\ket{+\omega_3} - \ket{-\omega_3}\right)\right).
\end{eqnarray}
where we have utilized the approximations of eigenstates (\ref{approx:uplus}) and (\ref{approx:uminus}) for large $N$. The state of the walk after $t$ steps is then given by
\begin{eqnarray}
	\nonumber \ket{\phi_\Omega(t)} &=& \hat{U}_{s,r}^t\ket{\Omega_s} = \frac{(-1)^t}{2}\ket{-1} + \\
    \nonumber & & +  \frac{i}{2\sqrt{2}}\left(e^{2i\omega_2 t}\ket{+\omega_1} - e^{- 2i\omega_2 t}\ket{-\omega_1}\right) + \\
    \nonumber & & + \frac{i}{2\sqrt{2}}\left(e^{ i\omega_2 t}\ket{+\omega_2} - e^{- i\omega_2 t}\ket{-\omega_2}\right) - \\
    \label{phi:omega} & & - i\frac{(-1)^t}{2\sqrt{2}}\left(e^{ -i\omega_2 t}\ket{+\omega_3} - e^{ i\omega_2 t}\ket{-\omega_3}\right).
\end{eqnarray}
For the amplitude of state transfer to $\ket{\Omega_r}$ we obtain
\begin{eqnarray}
\nonumber \braket{\Omega_r}{\phi_\Omega (t)} &=&  \frac{1}{4}\left[(-1)^t+\cos\left(2\omega_2 t\right)-\right. \\
\label{fidelity:EqualSup} & & \hspace{24pt} -\left.\cos\left(\omega_2 t\right)\left(1+(-1)^t\right)\right].
\end{eqnarray}
We see that there are rapid changes between odd and even steps. For odd number of steps $2t+1$ the amplitude turns into
\begin{eqnarray}
	\label{fidelity:EqualSup:odd} \braket{\Omega_r}{\phi_\Omega (2t+1)} = -\frac{1}{2}\sin^2{(\omega_2 (2t+1))},
\end{eqnarray}
so the fidelity can reach $\frac{1}{4}$ at most. For even number of steps $2t$ the amplitude (\ref{fidelity:EqualSup}) equals 
\begin{eqnarray}
\label{fidelity:EqualSup:even} \braket{\Omega_r}{\phi_\Omega (2t)} & = & 
-\cos{(2\omega_2 t)}\sin^2\left(\omega_2 t\right). 
\end{eqnarray}
Hence, for the closest even integer to
\begin{eqnarray}
T  = 2t \approx \frac{\pi}{\omega_2} \approx \pi \sqrt{\frac{N}{2}}+O\left(\frac{1}{\sqrt{N}}\right) .
\label{transfer:stepNumber}
\end{eqnarray}
the amplitude (\ref{fidelity:EqualSup:even}) is close to one, i. e. we obtain state transfer from $\ket{\Omega_s}$ to $\ket{\Omega_r}$ with high fidelity. We note that (\ref{fidelity:EqualSup:even}) results in the same fidelity as for the initial state (\ref{sr:init:old}) studied in \cite{stefanak2016} (see eq. (17) of the aforementioned paper).

\subsection{Transfer of loop states}
\label{sec:sta:loop}

In the case of transfer of the loop states the initial and target states are
\begin{eqnarray}
 \ket{s,s}  = \ket{\nu_1} , \quad  \ket{r,r} = \ket{\nu_2} .
\end{eqnarray}
In the eigenvector basis we find that the for large $N$ the loop states can be approximated by
\begin{eqnarray}
	\nonumber \ket{s,s} &\approx& \frac{1}{2}\ket{(1)_1} + \frac{1}{2\sqrt{2}}\ket{(1)_2} + \frac{1}{4}\left(\ket{+\omega_1}+\ket{-\omega_1}\right)+ \\
 \nonumber & &  + \frac{1}{2}\left(\ket{+\omega_2} + \ket{-\omega_2}\right), \\
\nonumber  \ket{r,r} &\approx& \frac{1}{2}\ket{(1)_1} + \frac{1}{2\sqrt{2}}\ket{(1)_2} + \frac{1}{4}\left(\ket{+\omega_1}+\ket{-\omega_1}\right) -\\
\label{loops:eigenvec} & &  -\frac{1}{2}\left(\ket{+\omega_2} + \ket{-\omega_2}\right).
\end{eqnarray}
The state of the walk after $2t$ steps equals
\begin{eqnarray}
	\nonumber \ket{\phi_l(2t)} &=& \hat{U}_{s,r}^{2t}\ket{s,s} \\
    \nonumber &=&  \frac{1}{2}\ket{(1)_1}+\frac{1}{2\sqrt{2}}\ket{(1)_2}+ \\
    \nonumber & & + \frac{1}{4}\left(e^{4 i \omega_2 t}\ket{+\omega_1}+e^{- 4 i \omega_2 t}\ket{-\omega_1}\right)+\\
    \label{phi:l} & & + \frac{1}{2}\left(e^{ 2i \omega_2 t}\ket{+\omega_2} + e^{- 2 i\omega_2 t}\ket{-\omega_2}\right).
\end{eqnarray}
The scalar product of $\ket{\phi_l(2t)}$ and the target state is then given by
\begin{eqnarray}
\label{fidelity3}     \braket{r,r}{\phi_l(2t)} &=& 
  \sin^4\left(\omega_2 t\right).
\end{eqnarray}
Hence, we achieve state transfer with high fidelity in the same number of steps as for the equal weight superposition states (\ref{transfer:stepNumber}).

\begin{figure}[h!]
\centering
\includegraphics[width=0.45\textwidth]{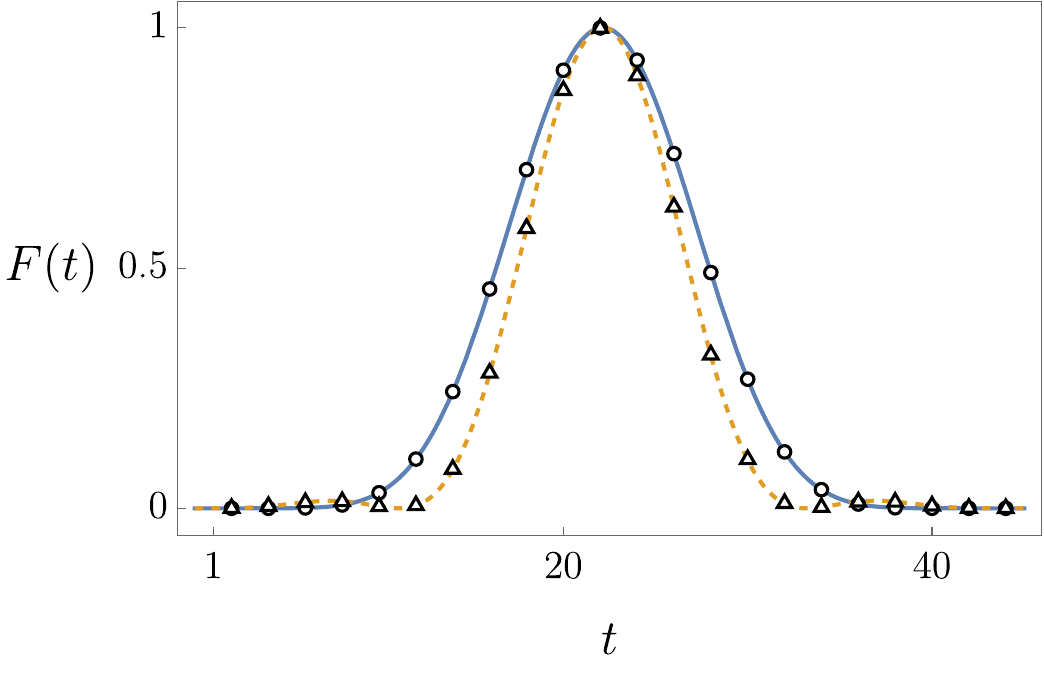}
\caption{Evolution of fidelity of state transfer from $\ket{s,s}$ to $\ket{r,r}$ (circles, full blue curve) and from $\ket{\Omega_s}$ to $\ket{\Omega_r}$ (triangles, dashed orange curve) on a graph with $100$ vertices. Only even time-steps are considered. Black data points are from numerical simulation, the curves are given by the square of the amplitudes (\ref{fidelity3}) and (\ref{fidelity:EqualSup:even}). Maximal fidelity of state transfer from $\ket{\Omega_s}$ to $\ket{\Omega_r}$ is reached at the same number of step as for the loops which is given by (\ref{transfer:stepNumber}) .}
\label{fig:sta:superpos}
\end{figure}

For comparison of state transfer of loops and equal weight superposition states see Fig. \ref{fig:sta:superpos}.

\subsection{Transfer of qubit state and finite-size effects of the graph}
\label{sec:sta:qubit}

Our results show that we can achieve state transfer of both $\ket{s,s}$ to $\ket{r,r}$ and $\ket{\Omega_s}$ to $\ket{\Omega_r}$ in the same run-time. Moreover, we can perform a state transfer of any superposition of these two orthonormal states. Hence, it is possible to transfer an arbitrary state of a qubit between two hubs in time given by the closest even integer to (\ref{transfer:stepNumber}). Consider a general qubit state at the sender encoded as
\begin{equation}
    \ket{\psi_s} = a \ket{s,s} + b\ket{\Omega_s} \equiv a\ket{0} + b \ket{1}.
\label{qubit:s}
\end{equation}
The target qubit state at receiver vertex is
\begin{equation}
\label{qubit:r}
    \ket{\psi_r} = a \ket{r,r} + b\ket{\Omega_r}.
\end{equation}
Time evolution of the initial state (\ref{qubit:s}) after $2t$ steps is given by
\begin{equation}
    \ket{\psi(2t)} = a\ket{\phi_l(2t)} + b \ket{\phi_\Omega(2t)}, 
\end{equation}
where $\ket{\phi_l(2t)}$ and $\ket{\phi_\Omega(2t)}$ are given by (\ref{phi:l}) and (\ref{phi:omega}). Amplitude of state transfer from (\ref{qubit:s}) to (\ref{qubit:r}) then reads
\begin{eqnarray}
\nonumber \braket{\psi_r}{\psi(2t)} & = & |a|^2\braket{r,r}{\phi_l(2t)} + |b|^2\braket{\Omega_r}{\phi_\Omega(2t)} + \\
\label{sta:ampl:qubit} & & + a\overline{b} \braket{\Omega_r}{\phi_l(2t)} + \overline{a} b \braket{r,r}{\phi_\Omega(2t)},
\end{eqnarray}
where $\braket{\Omega_r}{\phi_\Omega(2t)}$ and $\braket{r,r}{\phi_l(2t)}$ were determined in (\ref{fidelity:EqualSup}) and (\ref{fidelity3}). The remaining transition amplitudes are found to be
\begin{eqnarray}
\nonumber \braket{\Omega_r}{\phi_l(2t)} & = & \frac{1}{\sqrt{2}} \sin^2{\left(\omega_2 t\right)}\sin{\left(2\omega_2 t\right)} , \\
\braket{r,r}{\phi_\Omega(2t)} & = & - \braket{\Omega_r}{\phi_l(2t)},
\end{eqnarray}
which both vanish when the number of steps $T$ is taken as the closest even integer to (\ref{transfer:stepNumber}). Since $\braket{\Omega_r}{\phi_\Omega(T)}$ and $\braket{r,r}{\phi_l(T)}$ both tend to one for a large graph, we can transfer an arbitrary qubit state from the sender to the receiver vertex.

\begin{figure}[h]
    \centering
    \includegraphics[width=0.45\textwidth]{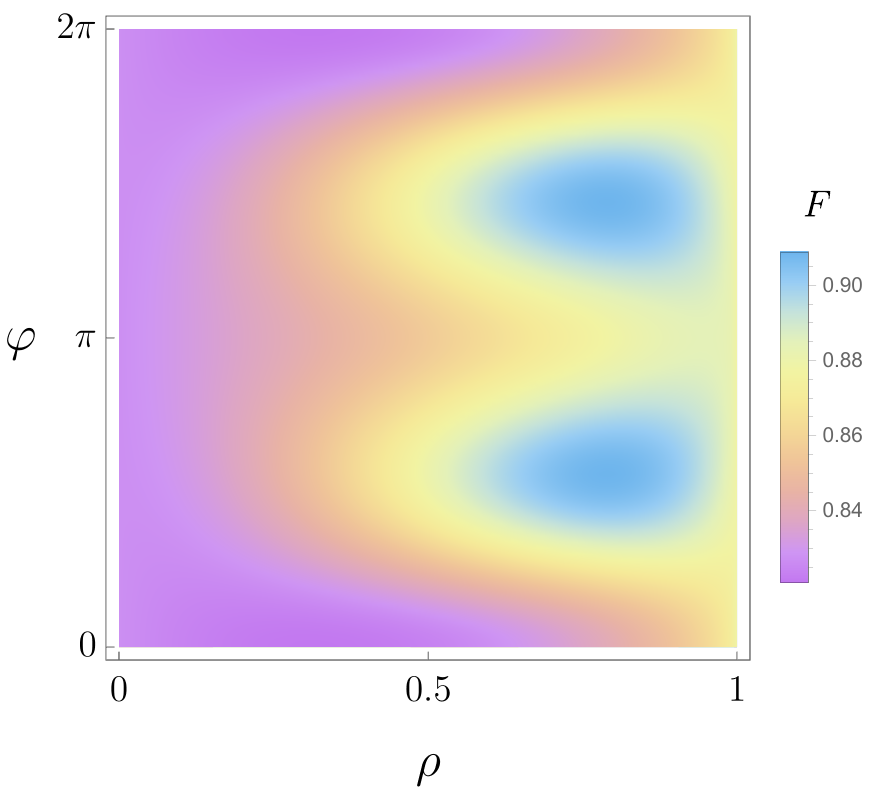}\vspace{12pt}
        \includegraphics[width=0.45\textwidth]{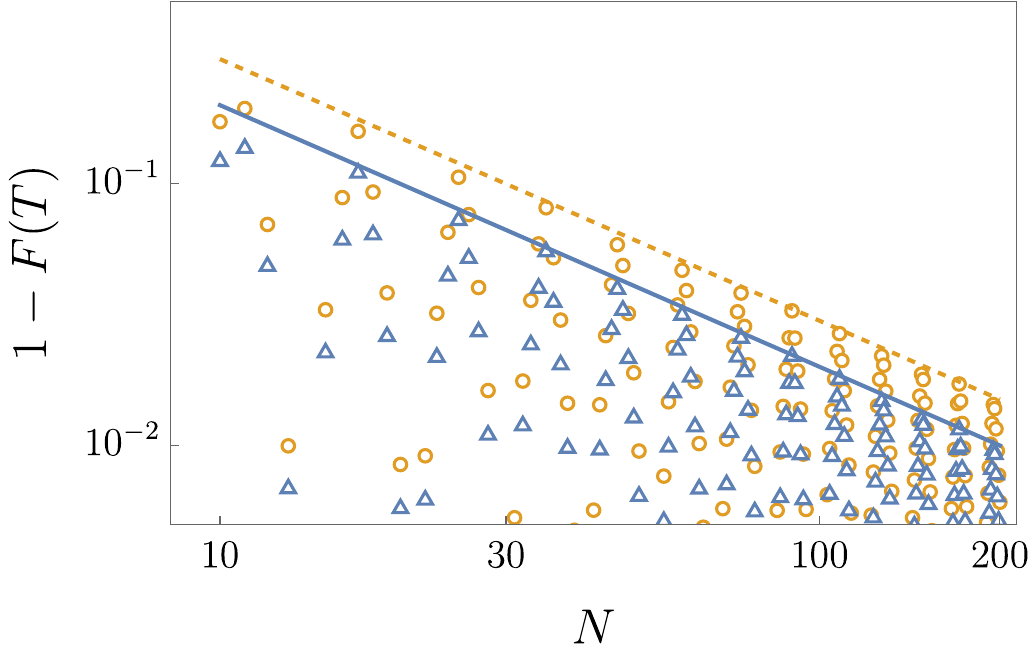}
    \caption{Upper plot: Fidelity of state transfer of a qubit (\ref{qubit:rho:phi}) as a function of $\rho$ and $\varphi$. The graph in consideration has $N=10$ vertices. The number of steps is chosen as the closest even integer to (\ref{transfer:stepNumber}), which for $N=10$ results in $T=6$. Lower plot: Convergence of fidelity of state transfer of the loop state $\ket{s,s}$ (blue triangles) and equal weight superposition state $\ket{\Omega_s}$ (yellow circles) as a function of size of the graph $N$. The plot is on log-log scale. The full blue line is given by the inverse power law $2/N$, for the dashed yellow line we have chosen $3/N$.}
    \label{fig:sta:qubit}
\end{figure}

We note that our result were derived utilizing the approximation of eigenstates (\ref{approx:uplus}), (\ref{approx:uminus}) for large graph, which neglected the $O\left(\frac{1}{\sqrt{N}}\right)$ terms in the eigenvectors. For smaller values of $N$ these contributions might be non-negligible and affect the fidelity of state transfer. In addition, the influence can be state dependent. To elucidate the finite-size effects we numerically investigated the fidelity of state transfer of a qubit state
\begin{equation}
\label{qubit:rho:phi}
    \ket{\psi} = \rho\ket{0} + e^{i\varphi}\sqrt{1-\rho^2}\ket{1},
\end{equation}
on a graph with 10 vertices. The results are displayed in the upper plot of Fig.~\ref{fig:sta:qubit}. For both parameters $\rho$ and $\varphi$ we split the corresponding interval range into 100 elements and evaluate the fidelity at the number of steps given by the closest even integer to (\ref{transfer:stepNumber}). The plot indicates that the worst fidelity is obtained for the state $\ket{1}$, i.e. $\ket{\Omega_s}$. On a graph of 10 vertices the fidelity ranges between 0.82 and 0.9.

In the lower plot of Fig.~\ref{fig:sta:qubit} we show the fidelity of state transfer for the basis states as a function of the graph size $N$. The plot is on the log-log scale to unravel the power-law behaviour $F(T) = 1- O\left(\frac{1}{N}\right)$. Note that the rapid drops and increases in the figure are due to the discreteness of the optimal time $T$, which has to be chosen as the closest even integer to (\ref{transfer:stepNumber}). The plot indicates that changing the qubit state to be transferred alters the pre-factor but not the power law dependence.

\subsection{Transfer of the connecting arcs}
\label{sec:sta:arcs}

Let us now consider the situation when the sender and the receiver know their positions. In such a case, they can transfer the states corresponding to the arcs on the connecting edge, i.e. the state $\ket{\nu_3}$ will be mapped onto $\ket{\nu_4}$. We show that this is achieved in the same runtime as for the loop and the equal weight superposition. 

The initial and the target states decomposed into the eigenvectors of $\hat U_{s,r}$ have the following form
\begin{eqnarray}
\nonumber \ket{\nu_3} & \approx &  -\frac{1}{2}\ket{(1)_1} + \frac{1}{2\sqrt{2}}\ket{(1)_2} + \frac{1}{4}(\ket{+\omega_1} + \ket{-\omega_1}) + \\
\nonumber & & +\frac{1}{2} (\ket{+\omega_3} + \ket{-\omega_3}), \\
\nonumber \ket{\nu_4} & \approx & -\frac{1}{2}\ket{(1)_1} + \frac{1}{2\sqrt{2}}\ket{(1)_2} + \frac{1}{4}(\ket{+\omega_1} + \ket{-\omega_1}) - \\
 & & - \frac{1}{2} (\ket{+\omega_3} + \ket{-\omega_3}).
\end{eqnarray}
In even steps $2t$, the state of the walk is given by
\begin{eqnarray}
\nonumber \ket{\phi_{sr}(2t)} & = & -\frac{1}{2}\ket{(1)_1} + \frac{1}{2\sqrt{2}}\ket{(1)_2} + \\
\nonumber & & + \frac{1}{4}(e^{4i \omega_2 t}\ket{+\omega_1} + e^{-4i \omega_2 t}\ket{-\omega_1}) + \\
& & + \frac{1}{2} (e^{-2i \omega_2 t}\ket{+\omega_3} + e^{2i \omega_2 t}\ket{-\omega_3}).
\end{eqnarray}
For the amplitude of state transfer in even steps we find
\begin{equation}
    \braket{\nu_4}{\phi_{sr}(2t)} = \sin^4{(\omega_2 t)},
\end{equation}
i.e., it behaves the same as for the transfer of loop states (\ref{fidelity3}). Hence, the arc from the sender to the receiver is transferred into its opposite in the runtime given by (\ref{transfer:stepNumber})

We note that the vectors $\ket{\nu_{3,4}}$ and $\ket{\Omega_{s,r}}$ are linearly independent but not mutually orthogonal. Nevertheless, from (\ref{omega:sr}) we find
\begin{equation}
    \braket{\nu_3}{\Omega_s} = \braket{\nu_4}{\Omega_r} = \frac{1}{\sqrt{N-1}} 
\end{equation}
so the overlap vanishes with increasing size of the graph $N$. Within the approximations we have used in our derivations, the states $\ket{s,s} = \ket{\nu_1}$, $\ket{\Omega_s}\approx \ket{\nu_5}$ and $\ket{\nu_3}$ are orthogonal. Their superposition represent a qutrit state, which can be transferred from the sender to the receiver vertex, provided that the two communicating parties know which edge connects them. Note that $\ket{\nu_1}$, $\ket{\nu_3}$ and $\ket{\nu_5}$ are the only basis states which are localized on the sender vertex. Hence, in this model we cannot transfer a higher dimensional qudit state with $d\geq 4$.

We have also tested numerically the effect of finite graph size on the fidelity of state transfer of the connecting arcs. The simulation reveals that for every $N$ the fidelity is exactly equal to the one for the state transfer of the loop state, which is depicted in the lower plot of Figure~\ref{fig:sta:qubit} with the blue triangles.

\section{State transfer between multiple sender and receiver hubs}
\label{sec:sta:multi}

We now expand the investigation of the previous Section by considering $S>1$ senders and $R>1$ receivers from sets ${\cal S}$ and ${\cal R}$, respectively, with $S+R=M$. We consider the same dynamics as for the search and state transfer between two marked vertices, i.e. the local loop weights are chosen as $l_v = N-d_v$ and on the marked hubs the coin is multiplied by -1. Having multiple senders and receivers leads to the following modification of the invariant subspace we have constructed in the previous Section for $S=R=1$. First, states (\ref{s:r:inv1}) corresponding to the loops and connecting arcs between senders and receivers are adjusted according to  
\begin{eqnarray}
\label{s:R:inv1} \ket{\nu_1} & = & \frac{1}{\sqrt{S}}\sum_{s\in{\cal S}}\ket{s,s}, \\
\nonumber \ket{\nu_2} & = & \frac{1}{\sqrt{R}} \sum\limits_{r\in {\cal R}}\ket{r,r} \\
\nonumber \ket{\nu_3} & = & \frac{1}{\sqrt{RS}} \sum_{s\in {\cal S}}\sum_{r\in {\cal R}} \ket{s,r} , \\
\nonumber \ket{\nu_4} & =  & \frac{1}{\sqrt{RS}} \sum_{s\in {\cal S}}\sum_{r\in {\cal R}}\ket{r,s} ,
\end{eqnarray}
where we have utilized the fact that evolution operator treats all senders and receivers equally and we can group them together. Next, states connecting sender and receiver vertices to the rest of the graph  (\ref{s:r:inv2}) are changed into 
\begin{eqnarray}
\label{s:R:inv2}    \ket{\nu_5} & = & \frac{1}{\sqrt{S(N-M)}} \sum_{s\in{\cal S}}\sum_{v \notin {\cal S}\cup{\cal R}}\ket{s,v} , \\
\nonumber \ket{\nu_6} & = & \frac{1}{\sqrt{R(N-M)}} \sum\limits_{r\in {\cal R}}\sum_{\substack{v \notin {\cal S}\cup {\cal R}}}\ket{r,v}, \\
\nonumber    \ket{\nu_7} & = &  \frac{1}{\sqrt{S(N-M)}} \sum_{s\in{\cal S}}\sum_{\substack{v\notin {\cal S}\cup {\cal R}}}\ket{v,s} , \\
\nonumber    \ket{\nu_8} & = & \frac{1}{\sqrt{R(N-M))}} \sum\limits_{r\in {\cal R}}\sum_{v \notin {\cal S}\cup {\cal R}}\ket{v,r} .
\end{eqnarray}
The state (\ref{s:r:inv3}) corresponding to the loops and arc between non-marked vertices is modified as follows
\begin{eqnarray}
\nonumber \ket{\nu_{9}} & = & \frac{1}{(N-M)} \sum_{v\notin {\cal S}\cup {\cal R}}\left(\frac{}{}\sqrt{N}\ket{\Omega_v(N-d_v)} - \right. \\
\label{s:R:inv3} & & \hspace{48pt} \left. - \sum\limits_{s\in {\cal S}}\ket{v,s} - \sum\limits_{r\in {\cal R}}\ket{v,r}\right).
\end{eqnarray}
To complete the invariant subspace we have to add two states describing the superpositions of arcs connecting sender hubs together, and the same for receiver hubs, i.e. 
\begin{eqnarray}
 \nonumber    \ket{\nu_{10}} & = & \frac{1}{\sqrt{S(S-1)}} \sum\limits_{s \neq s'\in {\cal S}} \ket{s,s'}, \\
   \ket{\nu_{11}} & = & \frac{1}{\sqrt{R(R-1)}} \sum\limits_{r \neq r'\in {\cal R}} \ket{r,r'},
  \label{s:R:inv4}  
\end{eqnarray}
 which are not present when we have only one sender and one receiver vertex.

We leave the detailed investigation of the evolution operator $\hat U_{{\cal S},{\cal R}}$
\begin{eqnarray}
\hat U_{{\cal S},{\cal R}} & = & \hat S\ \left(\bigoplus_{v\notin  {\cal S}, {\cal R}} \left(\hat G_v(N-d_v)\right) \right. \\
\nonumber & & \hspace{24pt} \left.\bigoplus_{s\in {\cal S}}   (-\hat{G}_s(1)) \bigoplus_{r\in{\cal R}}  (-\hat{G}_r(1))\right) .
\label{evol:op:SR}
\end{eqnarray}
for Appendix \ref{app:c}. The spectrum of $\hat U_{{\cal S},{\cal R}}$ has a similar form as for $S=R=1$, i.e. it consists of eigenvalues $\pm 1$ and three complex conjugated pairs $\lambda_{j}^{(\pm)} = e^{\pm i \omega_j}$, where the phases are given by
\begin{eqnarray}
\label{sta:multi:omega} \omega_1 & = & \arccos{\left(1-\frac{2(R+S)}{N}\right)}, \\
\nonumber \omega_2 & = & \frac{\omega_1}{2}, \quad \omega_3 = \pi - \omega_2 .
\end{eqnarray}
Eigenvalue 1 now has a degeneracy of 4. The corresponding eigenvectors are denoted by $\ket{(1)_m}$ , $m = 1,\ldots,4$, (see formulas (\ref{s:R:ev1})), $\ket{-1}$ (equation (\ref{s:R:evm1})) and $\ket{\pm\omega_j}$, $j=1,2,3$. Eigenvectors $\ket{\pm\omega_1}$ are given by (\ref{s:R:ev:omega1}), and approximations of $\ket{\pm\omega_2}$ and $\ket{\pm\omega_3}$ for large $N\gg R,S$ can be found in the formula (\ref{s:R:approx2}).

To investigate state transfer we consider evolution from two initial states supported on the sender vertices corresponding to the loops $(\ket{\nu_1})$ and the equal weight superposition of outgoing arcs $(\ket{\nu_5})$. We analyze the transition amplitudes to the states supported on the receiver vertices ($\ket{\nu_k}$, $k=2,4,6,11$). As we will see, for $S,R\neq 1$ it is not possible to transfer specific quantum states from the senders to the corresponding states on the receiver vertices. Nevertheless, if we relax the requirements and focus solely on transfer of the quantum walker from senders to receiver vertices irrespective of the internal state, we find that this task can be achieved in certain regimes of $S$ and $R$.

Let us begin with the superposition of loop states on the sender vertices $\ket{\nu_1}$. In this case the state of the walk after $t$ steps reads
\begin{eqnarray}
\ket{\phi_l(t)} & \approx & \frac{1}{M}\left(\frac{R}{\sqrt{S}} \ket{(1)_1} + \sqrt{\frac{S}{2}}\ket{(1)_2} + \right. \\
\nonumber & & \left. + M\sqrt{\frac{S-1}{S}}\ket{(1)_3} + \right. \\
\nonumber & & + \left.  \frac{\sqrt{2}}{2} \left[e^{i 2\omega_2 t}\ket{+\omega_1} + e^{-i 2\omega_2 t}\ket{-\omega_1}\right] + \right. \\
\nonumber & & + \left. \sqrt{R} \left[e^{i \omega_2 t} \ket{+\omega_2} + e^{-i \omega_2 t} \ket{-\omega_2}\right] \right) ,
\end{eqnarray}
within the approximations of (\ref{s:R:approx1}) and (\ref{s:R:approx2}) for $N\gg R,S$. Transition amplitudes to the states at the receiver vertices are given by
\begin{eqnarray}
 \braket{\nu_2}{\phi_l(t)} & = & \frac{4\sqrt{RS}}{M^2} \sin^4{\left(\frac{\omega_2 t}{2}\right)}, \\
 \nonumber  \braket{\nu_4}{\phi_l(t)} & = & -\frac{2\sqrt{R}}{M^2} (R + S\cos(\omega_2 t))\sin^2{\left(\frac{\omega_2 t}{2}\right)}, \\
 \nonumber \braket{\nu_6}{\phi_l(t)} & = &  -2\sqrt{\frac{RS}{M^3}} \sin(\omega_2 t) \sin^2{\left(\frac{\omega_2 t}{2}\right)}, \\
  \nonumber  \braket{\nu_{11}}{\phi_l(t)} & = &  \frac{4\sqrt{RS(R-1)}}{M^2} \sin^4{\left(\frac{\omega_2 t}{2}\right)}. 
\end{eqnarray}
We see that all amplitudes are vanishing with increasing number of sender and receiver vertices. The total probability of transfer to an arbitrary loop or arc originating from one of the receiver vertices is given by
\begin{equation}
\label{s:R:p1}
 P_l(t) = \sum_{k=2,4,6,11} |\braket{\nu_k}{\phi_l(t)}|^2 = \frac{4R}{(R+S)^2} \sin^4{\left(\frac{\omega_2 t}{2}\right)} .
\end{equation}
Hence, from the loop state $\ket{\nu_1}$ one can achieve unit transfer probability only in the case $S=R=1$ treated in the previous Section. Notice the asymmetry - for $R\gg S$ the maximal transfer probability decreases as $~R^{-1}$, while for $S\gg R$ the decrease is faster and follows $~S^{-2}$. 

Concerning the superposition of all outgoing arcs from all senders \begin{equation}
\label{init:s:R}
    \ket{\Omega_{\cal S}} = \sqrt{\frac{RS}{N-1}}\ket{\nu_3} + \sqrt{\frac{S(N-M)}{N-1}}\ket{\nu_5},
\end{equation}
we focus on a large graph with $N\gg R,S$ and approximate this initial state as $\ket{\nu_5}$. After $t$ steps the walk is described by the following vector 
\begin{eqnarray}
\ket{\phi_{\Omega}(t)} & \approx & \frac{1}{2\sqrt{R+S}}\left( (-1)^t \sqrt{2S} \ket{-1} + \right.\\
\nonumber & & \left. + i \sqrt{S}\left[e^{2i\omega_2 t} \ket{+\omega_1} - e^{-2i\omega_2 t} \ket{-\omega_1}\right] + \right. \\
\nonumber & & \left. + i\sqrt{R}\left[e^{i\omega_2 t} \ket{+\omega_2} - e^{-i\omega_2 t} \ket{-\omega_2}\right] -   \right. \\
\nonumber & & \left. - i(-1)^t\sqrt{R}\left[e^{-i\omega_2 t} \ket{+\omega_3} - e^{i\omega_2 t} \ket{-\omega_3}\right] \right) .
\end{eqnarray}
Turning to the transition amplitudes, we investigate odd and even steps separately. For even number of steps $2t$ we obtain
\begin{eqnarray}
 \nonumber \braket{\nu_2}{\phi_\Omega(2t)} & = &   -\braket{\nu_{6}}{\phi_l(2t)} , \\
  \nonumber \braket{\nu_4}{\phi_\Omega(2t)} & = & \sqrt{S}\braket{\nu_2}{\phi_\Omega(2t)} , \\
 \nonumber  \braket{\nu_6}{\phi_\Omega(2t)} & = & -\frac{2\sqrt{RS}}{M} \cos(2\omega_2 t)\sin^2(\omega_2 t), \\
 \braket{\nu_{11}}{\phi_\Omega(2t)} & = &  \sqrt{R-1}\braket{\nu_2}{\phi_\Omega(2t)} . 
\end{eqnarray}
The total transfer probability in even steps is given by 
\begin{eqnarray}
\label{sta:s:R:even}    
P_\Omega(2t) = \frac{4RS}{(R+S)^2} \sin^4{\left(\omega_2 t\right)},
\end{eqnarray}
which reaches the maximum
\begin{equation}
\label{pmax:even}
    P_{max}^{(even)} =  \frac{4RS}{(R+S)^2},
\end{equation}
for number of steps given by the closest even integer to
\begin{equation}
\label{sta:multi:T:even}
    T = 2t \approx \frac{\pi}{\omega_2} \approx \pi  \sqrt{\frac{N}{R+S}}.
\end{equation}
We point out that the result (\ref{pmax:even}) coincide with the one for the continuous time quantum walk (\ref{ctqw:fmax}).

Considering odd steps, the amplitudes at time $2t+1$ are equal to
\begin{eqnarray}
\nonumber     \braket{\nu_2}{\phi_\Omega(2t+1)} & = & 2\sqrt{\frac{RS}{M^3}} \sin(\omega_2 (2t+1)) \times \\
    \nonumber & & \times \sin^2{\left(\frac{\omega_2 (2t+1)}{2}\right)} , \\
\nonumber \braket{\nu_4}{\phi_\Omega(2t+1)} & = & -\sqrt{\frac{R}{M^3}}\left[R + S \cos{(\omega_2(2t+1))}\right] \times \\
\nonumber & & \times \sin(\omega_2 (2t+1)) , \\
\nonumber \braket{\nu_6}{\phi_\Omega(2t+1)} & = & -\frac{\sqrt{RS}}{M}\sin^2(\omega_2 (2t+1)) \\
\braket{\nu_{11}}{\phi_\Omega(2t+1)} & = & \sqrt{R-1}\braket{\nu_2}{\phi_\Omega(2t+1)} .
\end{eqnarray}
The total transfer probability in odd steps is then given by
\begin{eqnarray}
\label{sta:s:R:odd} 
P_\Omega(2t+1) & = & \frac{R}{R+S}\sin^2(\omega_2(2t+1)).
\end{eqnarray}
The maximum of value
\begin{equation}
\label{pmax:odd}
    P_{max}^{(odd)} = \frac{R}{R+S},
\end{equation}
is reached in a number of steps given by the closest odd integer to $T/2$.

\begin{figure}
    \centering
    \includegraphics[width=0.45\textwidth]{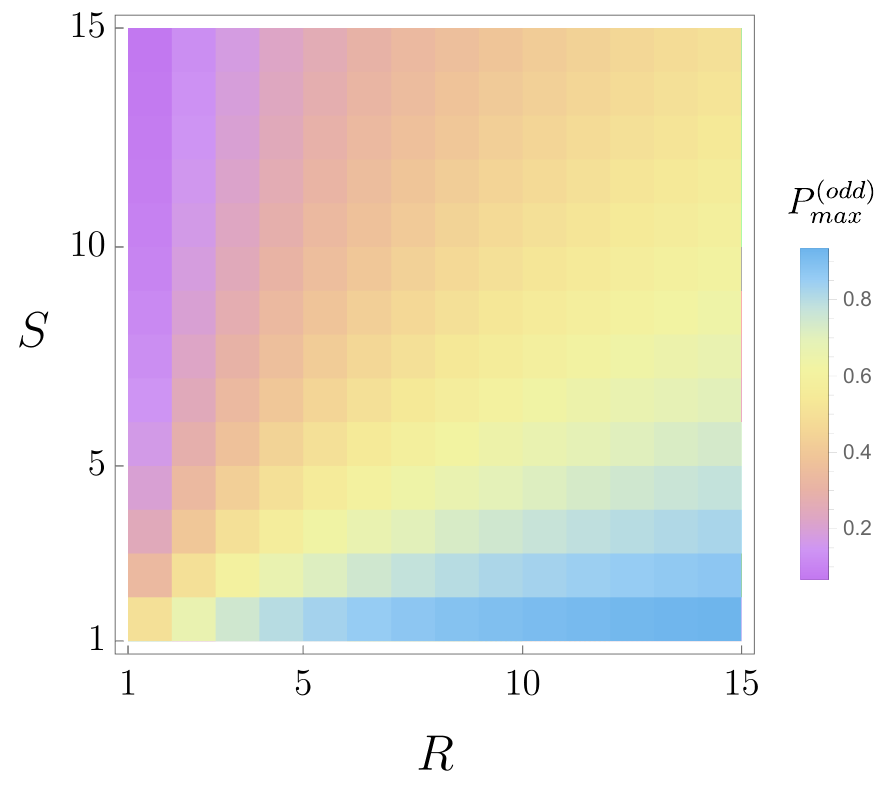}
    \caption{Maximal transfer probability from the state $\ket{\Omega_{\cal S}}$ in odd steps (\ref{pmax:odd}) as a function of $S$ and $R$.}
    \label{fig:transfer:max:sr}
\end{figure}

\begin{figure}
    \centering
    \includegraphics[width=0.45\textwidth]{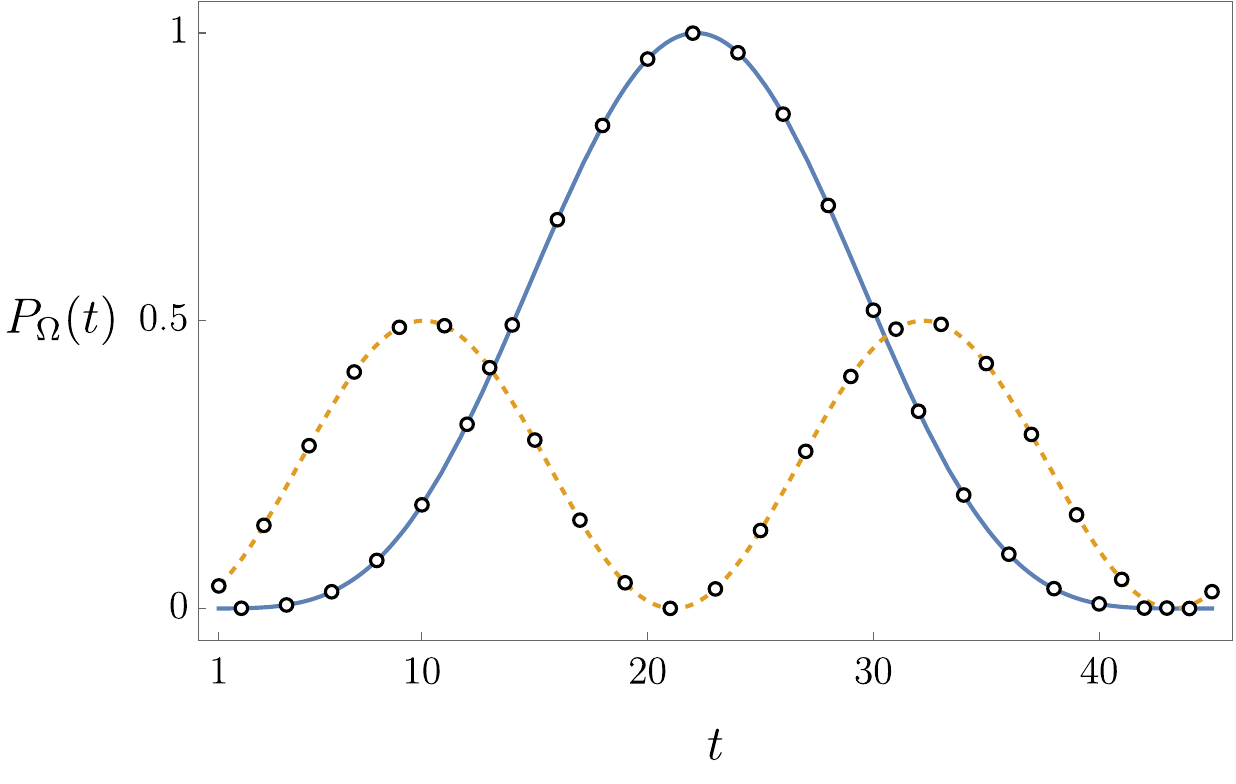}
    \caption{Transfer probability between 10 senders and 10 receivers on a graph with $N=1000$ vertices. As the initial state we choose $\ket{\Omega_{\cal S}}$. Circles correspond to the numerical simulation, the full blue curve captures even steps (\ref{sta:s:R:even}), the dashed yellow curve corresponds to odd steps (\ref{sta:s:R:odd}). Since $S=R$ we reach transfer probability close to 1 in a number of steps given by the closest even integer to (\ref{sta:multi:T:even}).}
    \label{fig:transfer:s=r}
\end{figure}

\begin{figure}
    \centering
    \includegraphics[width=0.45\textwidth]{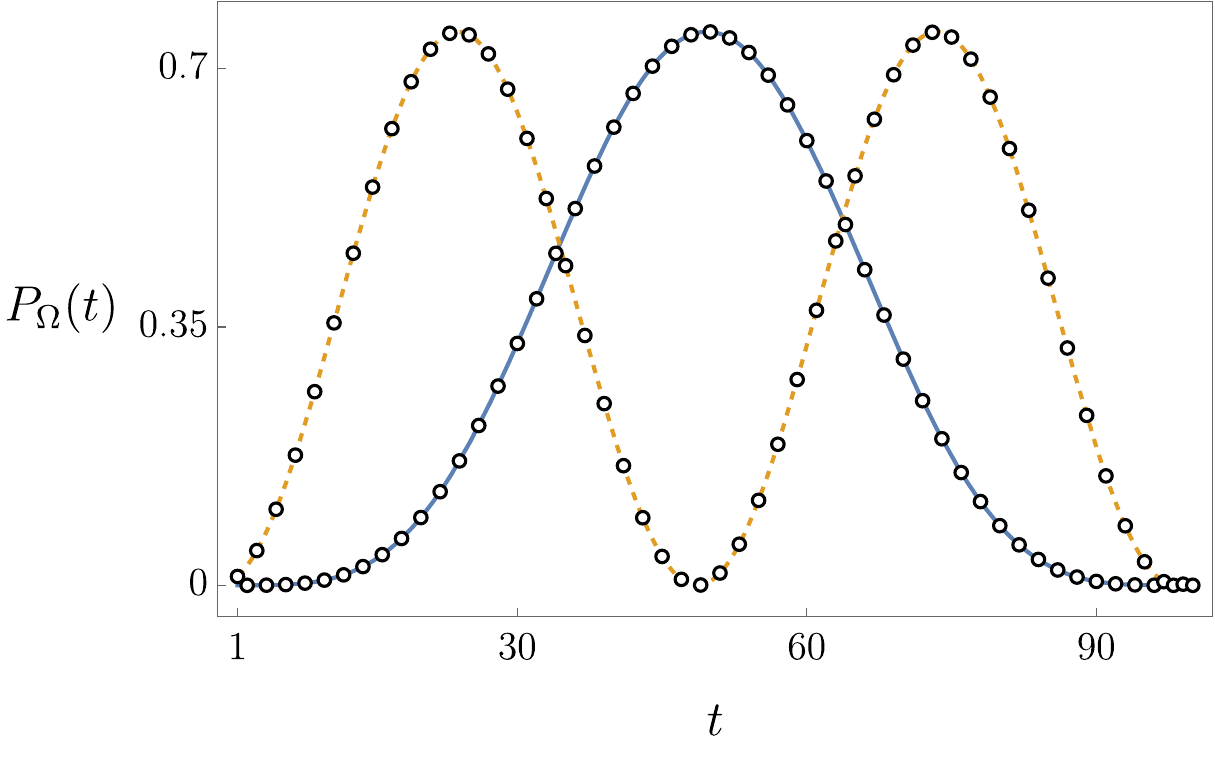}
    \includegraphics[width=0.45\textwidth]{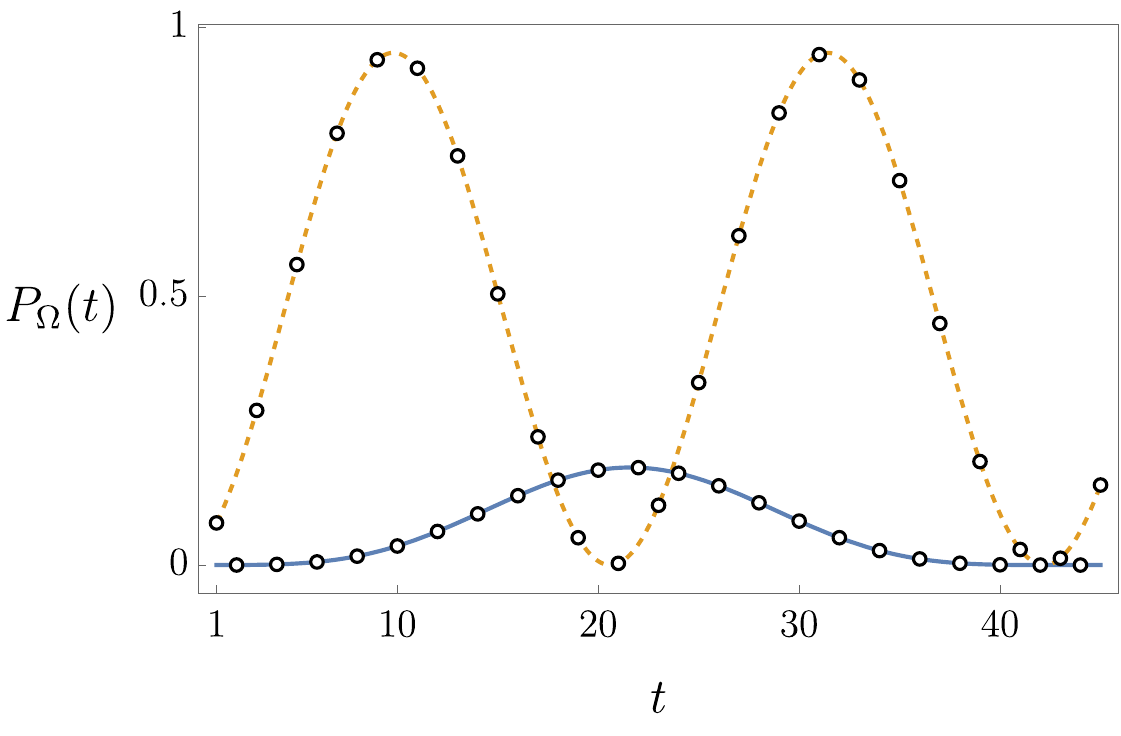}
    \caption{Transfer probability from a single sender vertex ($S=1$) to multiple receivers. We display the course of the transfer probability on a graph with $N=1000$ vertices with  $R=3$ and $R=20$ receivers for the initial state $\ket{\Omega_s}$ given in (\ref{init:s:R}) for $S=1$. The full blue curve capturing even steps is given by (\ref{sta:s:R:even}), the dashed yellow curve corresponding to odd steps follows from (\ref{sta:s:R:odd}). In the upper plot with $R=3$ the peaks for both even and odd steps reach value close to 0.75. In the lower plot corresponding to $R=20$ the peak for even steps is suppressed considerably, while we reach transfer probability close to 1 in a number of steps given by the closest odd integer to $T/2$, where $T$ can be found from (\ref{sta:multi:T:even}).}
    \label{fig:transfer:r}
\end{figure}

The derived results show that the discrete time quantum walk allows to transfer the walker initialized at the sender vertices in the state $\ket{\Omega_{\cal S}}$ to the receiver vertices with high probability in two regimes. The first regime occurs for $R\approx S$, which is the same as for the continuous time quantum walk. Here the probability in even steps reaches close to one (\ref{pmax:even}) for $T$ given by (\ref{sta:multi:T:even}). The second regime is given by $R\gg S$, which is absent for the continuous time case. In this case the transfer probability approaches one in odd steps (\ref{pmax:odd}) in half the runtime. We illustrate our results in Figures~\ref{fig:transfer:max:sr}, \ref{fig:transfer:s=r} and \ref{fig:transfer:r}.

In Figure~\ref{fig:transfer:max:sr} we show the maximal transfer probabilities in odd steps (\ref{pmax:odd}) as a function of the number of senders $S$ and the number of receivers $R$. For even steps, we refer to the continuous time case and Figure~\ref{fig:ctqw:transfer:max:sr}, since $P_{max}^{(even)}$ (\ref{pmax:even}) and $F_{max}$ (\ref{ctqw:fmax}) are equal.

Figure~\ref{fig:transfer:s=r} displays the course of the transfer probability for the case of equal number of senders and receivers. We see that the transfer probability is close to 1 for even number of steps $T$ given by (\ref{sta:multi:T:even}). In odd number of steps, the maximal value reached is 1/2 in approximately $T/2$ steps.

In Figure~\ref{fig:transfer:r} we consider transfer from a single sender vertex to multiple receivers. The two plots display the course of the transfer probability on a graph of $N=1000$ vertices with 3 receivers (upper plot) and 20 receivers (lower plot). We observe that for $R=3$ the peaks for even (\ref{sta:s:R:even}) and odd (\ref{sta:s:R:odd}) steps reach almost the same maximal value close to 0.75. For $R=20$ the peak for even steps is suppressed considerably, while in odd number of steps the transfer probability reaches close to 1.

\section{Conclusions}
\label{sec:concl}

Search and state transfer between multiple hubs on otherwise arbitrary graphs were investigated in detail. We have shown that both continuous-time and discrete-time quantum walks can be utilized for this purpose, since the models can be mapped to the complete graph with loops for a proper choice of the loop weights. In the continuous-time case the state transfer is achieved with high fidelity when the number of senders $S$ is close to the number of receivers $R$. The discrete-time walk has proven to be more versatile. Indeed, state transfer between multiple hubs is achieved also in the regime $R\gg S$, which is not possible in the continuous-time case. Moreover, when only a single sender and a single receiver is considered, the coin degree of freedom allows us to transfer multiple orthogonal states in the same run-time, expanding our earlier results \cite{stefanak2016}. We have shown that sender and receiver can always exchange an arbitrary qubit state, and in the case they know each others position this can be extended to a qutrit. This is not possible in the continuous time model, since there the walker does not have an internal degree of freedom, only the position.

There are several directions for generalizing our results. Our aim is to investigate if one can achieve state transfer from an ordinary vertex (i.e. not fully connected) to a hub. If the answer is positive we can perform state transfer between arbitrary vertices through the hub by switching the marked coin from the sender to the receiver after the transfer to the hub was completed, while keeping the hub vertex marked throughout the whole duration. Another approach to state transfer on arbitrary graphs is to utilize the framework based on the ergodic reversible Markov chains developed for both discrete-time \cite{ambainis_quadratic_2020} and continuous-time models \cite{apers_quadratic_2022}. This method was used to prove that quantum spatial search is quadratically faster over the classical search for an arbitrary number of marked vertices. We aim to modify this approach to investigate state transfer between multiple vertices which are not fully connected, where the reduction to a complete graph as in this paper is not possible.

\begin{acknowledgments}
M\v S acknowledges the financial support from Czech Grant Agency project number GA\v CR 23-07169S. SS is grateful for financial support from SGS22/181/OHK4/3T/14. 

\end{acknowledgments}

\appendix

\section{Evolution operator of the search}
\label{app:a}

\begin{widetext}
In this Appendix we treat the search evolution operator (\ref{dtqw:evol:search}) in the invariant subspace ${\cal I}$ spanned by vectors (\ref{basis:search1}), (\ref{basis:search2}) and (\ref{basis:search3}). The action of $\hat U_{\cal M}$ on the basis states is given by
\begin{eqnarray}
\nonumber  \hat U_{\cal M} \ket{\nu_1} & = & \left(1-\frac{2}{N}\right) \ket{\nu_1}  - \frac{2\sqrt{M-1}}{N} \ket{\nu_2}  -\frac{2\sqrt{N-M}}{N} \ket{\nu_4} , \\
\nonumber  \hat U_{\cal M} \ket{\nu_2} & = & - \frac{2\sqrt{M-1}}{N}\ket{\nu_1}  + \left(1 - 2\frac{M-1}{N}\right)\ket{\nu_2} - \frac{2\sqrt{(M-1)(N-M})}{N} \ket{\nu_4} , \\
\nonumber  \hat U_{\cal M} \ket{\nu_3} & = & -\frac{2\sqrt{N-M}}{N}\ket{\nu_1}  -\frac{2\sqrt{(M-1)(N-M})}{N}\ket{\nu_2} - \left(1-\frac{2M}{N}\right) \ket{\nu_4} ,\\
\nonumber  \hat U_{\cal M} \ket{\nu_4} & = & - \left(1-\frac{2M}{N}\right) \ket{\nu_3} + \frac{2\sqrt{M(N-M)}}{N} \ket{\nu_5} , \\
\label{evol:op:search} \hat U_{\cal M} \ket{\nu_5} & = & \frac{2\sqrt{M(N-M)}}{N}\ket{\nu_3} + \left(1-\frac{2M}{N}\right) \ket{\nu_5},
\end{eqnarray}
which shows that ${\cal I}$ is indeed invariant. 

Spectrum of the reduced evolution operator in the invariant subspace consists of eigenvalues $\pm 1$ and $e^{\pm i \omega}$ with the phase $\omega$ given in (\ref{eigenphase2}). We find that eigenvalue $1$ is doubly degenerate and the corresponding eigenvectors read
\begin{eqnarray}
\label{search:ev:p1} \ket{(1)_1} & = & \frac{1}{\sqrt{2 N M}} \left(\sqrt{N-M}\left(\ket{\nu_1} + \sqrt{M-1}\ket{\nu_2}\right) - M (\ket{\nu_3} + \ket{\nu_4}) - \sqrt{M(N-M)}\ket{\nu_5}\right), \\
\nonumber \ket{(1)_2} & = & \frac{1}{\sqrt{M}} \left(\sqrt{M-1}\ket{\nu_1} - \ket{\nu_2}\right).
\end{eqnarray}
Eigenvector corresponding to -1 is given by
\begin{eqnarray}
\label{search:ev:m1} \ket{-1} & = & \frac{1}{\sqrt{2N}} \left(\ket{\nu_1} + \sqrt{M-1}\ket{\nu_2}  - \sqrt{M}\ket{\nu_5} + \sqrt{N-M}\left(\ket{\nu_3}+\ket{\nu_4}\right) \right).
\end{eqnarray}
For the conjugate pair $\lambda_\pm$ we find the eigenvectors
\begin{eqnarray}
\label{search:ev:omega} \ket{\pm \omega} & = & \frac{1}{2}\left(\frac{1}{\sqrt{M}}\ket{\nu_1} + \sqrt{\frac{M-1}{M}}\ket{\nu_2} + \ket{\nu_5} \mp i(\ket{\nu_3}-\ket{\nu_4})\right).
\end{eqnarray}
Note that for a single marked hub the state $\ket{\nu_2}$ vanishes. In that case we have only 4-dimensional invariant subspace. Nevertheless, all results remain valid for $M=1$, except that eigenvalue 1 is simple as the state $\ket{(1)_2}$ equals zero.

\end{widetext}

\section{Evolution operator of state transfer between two hubs}
\label{app:b}

In this Appendix we investigate the evolution operator of the state transfer between two hubs (\ref{evol:op:2hubs}) in the invariant subspace spanned by vectors $\ket{\nu_j}$ given in equations (\ref{s:r:inv1}), (\ref{s:r:inv2}) and (\ref{s:r:inv3}). To simplify further calculations we utilize the invariance of $\hat U_{s,r}$ under exchange of sender and receiver vertices \cite{skoupy:2022}. This symmetry reduces the dimensionality of the problem, since we can decompose $\hat U_{s,r}$ into a direct sum of lower-dimensional operators. Consider the swap operator $\hat P$ which acts on the basis states $\ket{\nu_j}$ as
\begin{eqnarray}
	\nonumber \hat P \ket{\nu_1}&=&\ket{\nu_2},\quad  \hat P \ket{\nu_3} = \ket{\nu_4},\quad  \hat P \ket{\nu_5} = \ket{\nu_6},\\
	\hat P \ket{\nu_7}&=&\ket{\nu_8},\quad  	\hat P \ket{\nu_9} = \ket{\nu_9}.
\end{eqnarray}
Clearly, $\hat P$ commutes with $\hat{U}_{s,r}$ and therefore they have common eigenvectors. Since the square of $\hat P$ is the identity it has eigenvalues $\pm 1$. Hence, it divides the invariant subspace ${\cal I} = {\cal I}_{+}\oplus {\cal I}_{-} $ into a direct sum of symmetric states ${\cal I}_{+}$ and antisymmetric states ${\cal I}_{-}$. We find that ${\cal I}_{+}$ has dimension $5$ and ${\cal I}_{-}$ has dimension $4$. Subspace ${\cal I}_{+}$ is spanned by vectors 
\begin{eqnarray}
    \nonumber\ket{\sigma_1}&=&\frac{1}{\sqrt{2}}\left(\ket{\nu_1}+\ket{\nu_2}\right),\\
	\nonumber\ket{\sigma_2}&=&\frac{1}{\sqrt{2}}\left(\ket{\nu_3}+\ket{\nu_4}\right),\\
    \nonumber\ket{\sigma_3}&=&\frac{1}{\sqrt{2}}\left(\ket{\nu_5}+\ket{\nu_6}\right),\\
    \nonumber\ket{\sigma_4}&=&\frac{1}{\sqrt{2}}\left(\ket{\nu_7}+\ket{\nu_8}\right),\\ 
\ket{\sigma_5}&=&\ket{\nu_9}
\end{eqnarray}

Let us denote the restriction of the evolution operator $\hat U_{s,r}$ onto the symmetric subspace as $\hat U_+$. We find that it acts on the basis states $\ket{\sigma_j}$ as follows 
\begin{eqnarray}
    \nonumber\hat{U}_{+}\ket{\sigma_1}&=&\left(1-\frac{2}{N}\right)\ket{\sigma_1}-\frac{2}{N}\ket{\sigma_2}-\frac{2\sqrt{N-2}}{N}\ket{\sigma_4},\\
	\nonumber\hat{U}_{+}\ket{\sigma_2}&=&-\frac{2}{N}\ket{\sigma_1}+\left(1-\frac{2}{N}\right)\ket{\sigma_2}-\frac{2\sqrt{N-2}}{N}\ket{\sigma_4},\\
    \nonumber\hat{U}_{+}\ket{\sigma_3}&=&-\frac{2\sqrt{N-2}}{N}\ket{\sigma_1}-\frac{2\sqrt{N-2}}{N}\ket{\sigma_2} - \\
    \nonumber & & -\left(1-\frac{4}{N}\right)\ket{\sigma_4},\\
    \nonumber\hat{U}_{+}\ket{\sigma_4}&=& -\left(1-\frac{4}{N}\right) \ket{\sigma_3}+\frac{2\sqrt{2}\sqrt{N-2}}{N}\ket{\sigma_5,}\\
   \hat{U}_{+}\ket{\sigma_5}&=&\frac{2\sqrt{2}\sqrt{N-2}}{N}\ket{\sigma_3} + \left(1-\frac{4}{N}\right)\ket{\sigma_5} .
\end{eqnarray}
Spectrum of $\hat{U}_{+}$ is composed of eigenvalues $\pm 1$ and a conjugate pair $\lambda_{1}^{(\pm)}=e^{\pm i\omega_1}$ with eigenphase given by (\ref{sta:2:omega1}). Eigenvalue 1 has degeneracy 2, the corresponding eigenvectors are given by 
\begin{eqnarray}
\label{sta:ev:p1}
    \ket{(1)_1}&=&\frac{1}{\sqrt{2}}\left(\ket{\sigma_1}-\ket{\sigma_2}\right),\\
\nonumber	\ket{(1)_2}&=&\frac{1}{2\sqrt{N}}\left(\sqrt{N-2}\left(\ket{\sigma_1}+\ket{\sigma_2}\right) - \right. \\
 \nonumber  & & \qquad \qquad - \left. 
 2\left(\ket{\sigma_3}+\ket{\sigma_4}\right) - \sqrt{2}\sqrt{N-2}\ket{\sigma_5}\right).
\end{eqnarray}
Eigenvector corresponding to $-1$ reads
\begin{eqnarray}
	\nonumber\ket{-1}&=&\frac{1}{\sqrt{2N}}\left(\ket{\sigma_1}+\ket{\sigma_2}+\sqrt{N-2}\left(\ket{\sigma_3}+\ket{\sigma_4}\right) - \right.\\
\label{sta:ev:m1} & & \qquad \qquad \left. -\sqrt{2}\ket{\sigma_5}\right) .
\end{eqnarray}
For large $N$ the eigenvectors $\ket{(1)_2}$ and $\ket{-1}$ can be approximated by neglecting the $O\left(\frac{1}{\sqrt{N}}\right)$ terms with
\begin{eqnarray}
\nonumber \ket{(1)_2}  & \approx & \frac{1}{2}\left(\ket{\sigma_1}+\ket{\sigma_2}\right) - \frac{1}{\sqrt{2}}\ket{\sigma_5} , \\
\label{approx:uplus} \ket{-1} & \approx & \frac{1}{\sqrt{2}}\left(\ket{\sigma_3}+\ket{\sigma_4}\right). 
\end{eqnarray}
Concerning the conjugated pair we find that the corresponding  eigenvectors read
\begin{eqnarray}
    \nonumber \ket{\pm\omega_1} &=& \frac{1}{2}\left(\frac{1}{\sqrt{2}}\left(\ket{\sigma_1} + \ket{\sigma_2}\right) + \ket{\sigma_5}  \pm i(\ket{\sigma_4} - \ket{\sigma_3})\right).\\
    \label{sta:ev:omega1}
   \end{eqnarray}
   
Lets now move to the antisymmetric subspace ${\cal I}_{-}$ which is spanned by vectors 
\begin{eqnarray}
    \nonumber\ket{\tau_1}&=&\frac{1}{\sqrt{2}}\left(\ket{\nu_1}-\ket{\nu_2}\right),\\
	\nonumber\ket{\tau_2}&=&\frac{1}{\sqrt{2}}\left(\ket{\nu_3}-\ket{\nu_4}\right),\\
    \nonumber\ket{\tau_3}&=&\frac{1}{\sqrt{2}}\left(\ket{\nu_5}-\ket{\nu_6}\right),\\
   \ket{\tau_4}&=&\frac{1}{\sqrt{2}}\left(\ket{\nu_7}-\ket{\nu_8}\right).
\end{eqnarray}
We denote by $\hat{U}_{-}$ the restriction of $\hat U_{s,r}$ to ${\cal I}_{-}$, which acts as follows
\begin{eqnarray}
    \nonumber\hat{U}_{-}\ket{\tau_1} &=& \left(1-\frac{2}{N}\right) \ket{\tau_1} + \frac{2}{N}\ket{\tau_2 } -\frac{2\sqrt{N-2}}{N}\ket{\tau_4},\\
	\nonumber\hat{U}_{-}\ket{\tau_2}&=&-\frac{2}{N}\ket{\tau_1} -\left(1-\frac{2}{N}\right)\ket{\tau_2}-\frac{2\sqrt{N-2}}{N}\ket{\tau_4},\\
    \nonumber\hat{U}_{-}\ket{\tau_3}&=&-\frac{2\sqrt{N-2}}{N}\ket{\tau_1}+\frac{2\sqrt{N-2}}{N}\ket{\tau_2} - \\
    \nonumber & & - \left(1-\frac{4}{N}\right)\ket{\tau_4},\\
  \hat{U}_{-}\ket{\tau_4}&=&-\ket{\tau_3}.
\end{eqnarray}
We find that $\hat U_-$ has two pairs of conjugated eigenvalues $\lambda_{2}^{(\pm)}=e^{\pm i\omega_2}$ and $\lambda_{3}^{(\pm)}=e^{\pm i\omega_3}$ with eigenphases given by equations (\ref{sta:2:omega23}). The corresponding eigenvectors read
\begin{eqnarray}
\label{sta:ev:omega2} \ket{\pm\omega_2} & = &   \frac{\cos{\left(\frac{\omega_2}{2}\right)}}{\sqrt{2}}\ket{\tau_1} \mp i \frac{\sin\left({\frac{\omega_2}{2}}\right)}{\sqrt{2}}\ket{\tau_2} \pm \\
    \nonumber & &  \pm \frac{i}{2} \left(e^{\pm i \frac{\omega_2}{2}}\ket{\tau_4} - e^{\mp i \frac{\omega_2}{2}}\ket{\tau_3}\right),\\
\label{sta:ev:omega3} \ket{\pm\omega_3} &=& \pm i \frac{\sin\left({\frac{\omega_2}{2}}\right)}{\sqrt{2}}\ket{\tau_1} + \frac{\cos{\left(\frac{\omega_2}{2}\right)}}{\sqrt{2}}\ket{\tau_2} \pm \\
    \nonumber & &  \pm \frac{i}{2} \left( e^{\pm i \frac{\omega_2}{2}}\ket{\tau_3} + e^{\mp i \frac{\omega_2}{2}}\ket{\tau_4} \right).
\end{eqnarray}
For large $N$ the phase $\omega_2$ tends to zero as $\sqrt{\frac{2}{N}}$. Omitting the $O\left(\frac{1}{\sqrt{N}}\right)$ terms the eigenvectors can be approximated by
\begin{eqnarray}
    \nonumber\ket{\pm\omega_2} & \approx &  \frac{1}{\sqrt{2}}\ket{\tau_1}  \pm  \frac{i}{2}\left(\ket{\tau_4} - \ket{\tau_3}\right) ,\\
    \label{approx:uminus}  \ket{\pm\omega_3} & \approx &  \frac{1}{\sqrt{2}}\ket{\tau_2} \pm  \frac{i}{2} \left(\ket{\tau_3} + \ket{\tau_4} \right) .
\end{eqnarray}

\section{Evolution operator of state transfer between multiple hubs}
\label{app:c}

In this Appendix we present the analysis of the evolution operator $\hat U_{{\cal S},{\cal R}}$ for state transfer between multiple senders and receivers (\ref{evol:op:SR}). We find that $\hat U_{{\cal S},{\cal R}}$ acts on the basis vectors of the invariant subspace (\ref{s:R:inv1}), (\ref{s:R:inv2}), (\ref{s:R:inv3}) and (\ref{s:R:inv4}) according to
\begin{widetext}
\begin{eqnarray}
\nonumber     \hat U_{{\cal S},{\cal R}} \ket{\nu_1} & = & \left(1-\frac{2}{N}\right)\ket{\nu_1} - \frac{2\sqrt{R}}{N}\ket{\nu_4} -    \frac{2 \sqrt{N-M}}{N} \ket{\nu_7} - \frac{2\sqrt{S-1}}{N} \ket{\nu_{10}}, \\
\nonumber     \hat U_{{\cal S},{\cal R}} \ket{\nu_2} & = & \left(1-\frac{2}{N}\right) \ket{\nu_2} -\frac{2\sqrt{S}}{N}\ket{\nu_3}  - \frac{2 \sqrt{N-M}}{N} \ket{\nu_8} -    \frac{2 \sqrt{R-1}}{N} \ket{\nu_{11}}, \\
\nonumber     \hat U_{{\cal S},{\cal R}} \ket{\nu_3} & = & - \frac{2\sqrt{R}}{N}\ket{\nu_1} + \left(1-\frac{2R}{N}\right)\ket{\nu_4}  -\frac{2 \sqrt{R(N-M)}}{N} \ket{\nu_7} - \frac{2 \sqrt{R(S-1)}}{N} \ket{\nu_{10}}, \\
\nonumber     \hat U_{{\cal S},{\cal R}} \ket{\nu_4} & = &  -\frac{2\sqrt{S}}{N} \ket{\nu_2} + \left(1-\frac{2}{N}\right)\ket{\nu_3}  - \frac{2 \sqrt{S(N-M)}}{N} \ket{\nu_8} -    \frac{2 \sqrt{S(R-1)}}{N} \ket{\nu_{11}} , \\
\nonumber     \hat U_{{\cal S},{\cal R}} \ket{\nu_5} & = & -\frac{2 \sqrt{N-M}}{N}\left(\ket{\nu_1} + \sqrt{R}\ket{\nu_4} +  \sqrt{S-1} \ket{\nu_{10}}\right)   -    \left(1-\frac{2M}{N}\right) \ket{\nu_7} , \\
\nonumber     \hat U_{{\cal S},{\cal R}} \ket{\nu_6} & = & - \frac{2 \sqrt{N-M}}{N}\left(\ket{\nu_2} + \sqrt{S}\ket{\nu_3}  +    \sqrt{R-1}\ket{\nu_{11}}\right) -\left(1-\frac{2M}{N}\right) \ket{\nu_8}, \\
\nonumber     \hat U_{{\cal S},{\cal R}} \ket{\nu_7} & = & -\left(1-\frac{2S}{N}\right)\ket{\nu_5} - \frac{2\sqrt{RS}}{N}\ket{\nu_6} -    \frac{2 \sqrt{S(N-M)}}{N} \ket{\nu_9}, \\
\nonumber     \hat U_{{\cal S},{\cal R}} \ket{\nu_8} & = &  \frac{2\sqrt{RS}}{N}\ket{\nu_5} - \left(1-\frac{2R}{N}\right)\ket{\nu_6} -    \frac{2 \sqrt{R(N-M)}}{N} \ket{\nu_9}, \\
\nonumber   \hat U_{{\cal S},{\cal R}} \ket{\nu_9} & = & \frac{2 \sqrt{N-M}}{N}\left(\sqrt{S}\ket{\nu_5} + \sqrt{R} \ket{\nu_6} \right) +    \left(1-\frac{2M}{N}\right)\ket{\nu_9}, \\
\nonumber \hat U_{{\cal S},{\cal R}} \ket{\nu_{10}} & = & - \frac{2 \sqrt{S-1}}{N} \left(\ket{\nu_1} + \sqrt{R}\ket{\nu_4} + \sqrt{N-M} \ket{\nu_7}\right)  +     \left(1-\frac{2(S-1)}{N}\right)\ket{\nu_{10}}, \\ 
\label{usR}  \hat U_{{\cal S},{\cal R}} \ket{\nu_{11}} & = & - \frac{2 \sqrt{R-1}}{N} \left(\ket{\nu_2} + \sqrt{S}\ket{\nu_3} + \sqrt{N-M} \ket{\nu_8}\right)  +     \left(1-\frac{2(R-1)}{N}\right)\ket{\nu_{11}} .
\end{eqnarray}
For $S\neq R$ the splitting of the invariant subspace into the symmetric and antisymmetric subspaces is no longer possible. Nevertheless, the evolution operator $\hat U_{{\cal S},{\cal R}}$ can be diagonalized analytically. The eigenvalues are again $\pm 1$ and three conjugated pairs $\lambda_j = e^{\pm i \omega_j}$, $j=1,2,3$,  where the phases $\omega_j$ are given in (\ref{sta:multi:omega}). Eigenvalue 1 has a degeneracy of 4, one can choose the basis in the degenerate subspace e.g. according to 
\begin{eqnarray}
\nonumber     \ket{(1)_1} & = & \frac{1}{M} \left(\frac{R}{\sqrt{S}} \ket{\nu_1} + \frac{S}{\sqrt{R}} \ket{\nu_2} - \sqrt{RS}\left(\ket{\nu_3} 
+ \ket{\nu_4}\right) + R \sqrt{\frac{S-1}{S}}\ket{\nu_{10}} + S \sqrt{\frac{R-1}{R}}\ket{\nu_{11}} \right) ,\\
\nonumber \ket{(1)_2} & = & \frac{1}{\sqrt{2N}M} \left[\sqrt{N-M}\left(\sqrt{S}\ket{\nu_1} + \sqrt{R}\ket{\nu_2} + \sqrt{RS}(\ket{\nu_3} + \ket{\nu_4}) - M\ket{\nu_9} + \right. \right.  \\
\nonumber & & \hspace{66pt} \left. \left. + \sqrt{S(S-1)} \ket{\nu_{10}} + \sqrt{R(R-1)} \ket{\nu_{11}} \right) -  M\left(\sqrt{S}(\ket{\nu_5} + \ket{\nu_7}) + \sqrt{R} (\ket{\nu_6} + \ket{\nu_8}) \right)\right], \\
 \nonumber  \ket{(1)_3} & = & \frac{1}{\sqrt{S}} (\sqrt{S-1}\ket{\nu_1} - \ket{\nu_{10}}), \\
 \label{s:R:ev1}  \ket{(1)_4} & = & \frac{1}{\sqrt{R}} (\sqrt{R-1}\ket{\nu_2} - \ket{\nu_{11}}).
\end{eqnarray}
The eigenvector corresponding to -1 reads
\begin{eqnarray}
\nonumber \ket{-1} & = & \frac{1}{\sqrt{2NM}} \left(\sqrt{S}\ket{\nu_1}  + \sqrt{R} \ket{\nu_2} + \sqrt{RS}(\ket{\nu_3} + \ket{\nu_4}) + \sqrt{N-M}\left[\sqrt{S}(\ket{\nu_5} + \ket{\nu_7}) + \sqrt{R}(\ket{\nu_6} + \ket{\nu_8})\right] - \right. \\
 \label{s:R:evm1}  & & \hspace{66pt} \left.  - M\ket{\nu_9} + \sqrt{S(S-1)}\ket{\nu_{10}} + \sqrt{R(R-1)}\ket{\nu_{11}}\right).
\end{eqnarray}
In the limit $N\gg R,S$ we can approximate the eigenstates $\ket{(1)_2}$ and $\ket{-1}$ with
\begin{eqnarray}
    \nonumber \ket{(1)_2} & \approx & \frac{1}{\sqrt{2}M} \left(\sqrt{S}\ket{\nu_1} + \sqrt{R}\ket{\nu_2} + \sqrt{RS}(\ket{\nu_3} + \ket{\nu_4}) - M\ket{\nu_9} + \sqrt{S(S-1)} \ket{\nu_{10}} +  \sqrt{R(R-1)} \ket{\nu_{11}} \right), \\
\label{s:R:approx1}    \ket{-1} & \approx & \frac{1}{\sqrt{2M}} \left[\sqrt{S}(\ket{\nu_5} + \ket{\nu_7}) + \sqrt{R}(\ket{\nu_6} + \ket{\nu_8}) \right].
\end{eqnarray}
For the complex conjugated pair $\lambda_{1}^{(\pm)}$ we find the following
\begin{eqnarray}
\nonumber    \ket{\pm\omega_1} & = & \frac{1}{2M} \left(\sqrt{S}\ket{\nu_1}  + \sqrt{R}\ket{\nu_2} + \sqrt{RS}(\ket{\nu_3} + \ket{\nu_4})  + M\ket{\nu_9} + \sqrt{S(S-1)} \ket{\nu_{10}} + \sqrt{R(R-1)} \ket{\nu_{11}} \pm \right. \\
 \label{s:R:ev:omega1} & & \hspace{60pt} \left. \pm i\sqrt{M} \left[\sqrt{S}(\ket{\nu_7} - \ket{\nu_5}) + \sqrt{R}(\ket{\nu_8} - \ket{\nu_6})\right]\right) .
\end{eqnarray}
The exact form of the eigenvectors corresponding to $\lambda_{2,3}^{(\pm)}$ is rather complicated, however, in the limit of $N \gg R,S$ they can be approximated by
\begin{eqnarray}
\nonumber \ket{\pm\omega_2} & \approx & \frac{1}{2M}\left(2\sqrt{R}\ket{\nu_1} - 2\sqrt{S}\ket{\nu_2} + (R-S)(\ket{\nu_3} + \ket{\nu_4}) + 2\sqrt{R(S-1)}\ket{\nu_{10}}  -2\sqrt{S(R-1)}\ket{\nu_{11}} \pm \right. \\
\nonumber & & \hspace{60pt} \left. \pm i\sqrt{M}\left[\sqrt{R}(\ket{\nu_7} -\ket{\nu_5} + \sqrt{S}(\ket{\nu_6} - \ket{\nu_8}))\right]  \right), \\
\label{s:R:approx2} \ket{\pm \omega_3} & \approx & \frac{1}{2M}\left(M(\ket{\nu_3} - \ket{\nu_4}) \pm i \sqrt{M}\left[\sqrt{R}(\ket{\nu_5} + \ket{\nu_7}) - \sqrt{S}(\ket{\nu_6} + \ket{\nu_8})\right] \right)
\end{eqnarray}
We point out that for $S=R=1$ the spectrum and the eigenvectors reduce to those derived in the Appendix \ref{app:b}. Note that in this case the states $\ket{\nu_{10}}$ and $\ket{\nu_{11}}$ need to be eliminated, since there no connecting edges when we have only a single sender and a single receiver.

\end{widetext}

\bibliography{biblio}

\begin{thebibliography}{74}%
\makeatletter
\providecommand \@ifxundefined [1]{%
 \@ifx{#1\undefined}
}%
\providecommand \@ifnum [1]{%
 \ifnum #1\expandafter \@firstoftwo
 \else \expandafter \@secondoftwo
 \fi
}%
\providecommand \@ifx [1]{%
 \ifx #1\expandafter \@firstoftwo
 \else \expandafter \@secondoftwo
 \fi
}%
\providecommand \natexlab [1]{#1}%
\providecommand \enquote  [1]{``#1''}%
\providecommand \bibnamefont  [1]{#1}%
\providecommand \bibfnamefont [1]{#1}%
\providecommand \citenamefont [1]{#1}%
\providecommand \href@noop [0]{\@secondoftwo}%
\providecommand \href [0]{\begingroup \@sanitize@url \@href}%
\providecommand \@href[1]{\@@startlink{#1}\@@href}%
\providecommand \@@href[1]{\endgroup#1\@@endlink}%
\providecommand \@sanitize@url [0]{\catcode `\\12\catcode `\$12\catcode
  `\&12\catcode `\#12\catcode `\^12\catcode `\_12\catcode `\%12\relax}%
\providecommand \@@startlink[1]{}%
\providecommand \@@endlink[0]{}%
\providecommand \url  [0]{\begingroup\@sanitize@url \@url }%
\providecommand \@url [1]{\endgroup\@href {#1}{\urlprefix }}%
\providecommand \urlprefix  [0]{URL }%
\providecommand \Eprint [0]{\href }%
\providecommand \doibase [0]{https://doi.org/}%
\providecommand \selectlanguage [0]{\@gobble}%
\providecommand \bibinfo  [0]{\@secondoftwo}%
\providecommand \bibfield  [0]{\@secondoftwo}%
\providecommand \translation [1]{[#1]}%
\providecommand \BibitemOpen [0]{}%
\providecommand \bibitemStop [0]{}%
\providecommand \bibitemNoStop [0]{.\EOS\space}%
\providecommand \EOS [0]{\spacefactor3000\relax}%
\providecommand \BibitemShut  [1]{\csname bibitem#1\endcsname}%
\let\auto@bib@innerbib\@empty
\bibitem [{\citenamefont {Aharonov}\ \emph {et~al.}(1993)\citenamefont
  {Aharonov}, \citenamefont {Davidovich},\ and\ \citenamefont
  {Zagury}}]{Aharonov1993}%
  \BibitemOpen
  \bibfield  {author} {\bibinfo {author} {\bibfnamefont {Y.}~\bibnamefont
  {Aharonov}}, \bibinfo {author} {\bibfnamefont {L.}~\bibnamefont
  {Davidovich}},\ and\ \bibinfo {author} {\bibfnamefont {N.}~\bibnamefont
  {Zagury}},\ }\bibfield  {title} {\bibinfo {title} {Quantum random walks},\
  }\href {https://doi.org/10.1103/PhysRevA.48.1687} {\bibfield  {journal}
  {\bibinfo  {journal} {Phys. Rev. A}\ }\textbf {\bibinfo {volume} {48}},\
  \bibinfo {pages} {1687} (\bibinfo {year} {1993})}\BibitemShut {NoStop}%
\bibitem [{\citenamefont {Meyer}(1996)}]{Meyer1996}%
  \BibitemOpen
  \bibfield  {author} {\bibinfo {author} {\bibfnamefont {D.~A.}\ \bibnamefont
  {Meyer}},\ }\bibfield  {title} {\bibinfo {title} {From quantum cellular
  automata to quantum lattice gases},\ }\href
  {https://doi.org/10.1007/BF02199356} {\bibfield  {journal} {\bibinfo
  {journal} {J. Stat. Phys.}\ }\textbf {\bibinfo {volume} {85}},\ \bibinfo
  {pages} {551} (\bibinfo {year} {1996})}\BibitemShut {NoStop}%
\bibitem [{\citenamefont {Farhi}\ and\ \citenamefont
  {Gutmann}(1998)}]{Farhi1998}%
  \BibitemOpen
  \bibfield  {author} {\bibinfo {author} {\bibfnamefont {E.}~\bibnamefont
  {Farhi}}\ and\ \bibinfo {author} {\bibfnamefont {S.}~\bibnamefont
  {Gutmann}},\ }\bibfield  {title} {\bibinfo {title} {Quantum computation and
  decision trees},\ }\href {https://doi.org/10.1103/PhysRevA.58.915} {\bibfield
   {journal} {\bibinfo  {journal} {Phys. Rev. A}\ }\textbf {\bibinfo {volume}
  {58}},\ \bibinfo {pages} {915} (\bibinfo {year} {1998})}\BibitemShut
  {NoStop}%
\bibitem [{\citenamefont {Aaronson}\ and\ \citenamefont
  {Ambainis}(2003)}]{aaronson_quantum_2003}%
  \BibitemOpen
  \bibfield  {author} {\bibinfo {author} {\bibfnamefont {S.}~\bibnamefont
  {Aaronson}}\ and\ \bibinfo {author} {\bibfnamefont {A.}~\bibnamefont
  {Ambainis}},\ }\bibfield  {title} {\bibinfo {title} {Quantum search of
  spatial regions},\ }in\ \href@noop {} {\emph {\bibinfo {booktitle} {44th
  {Annual} {IEEE} {Symposium} on {Foundations} of {Computer} {Science},
  {Proceedings}}}}\ (\bibinfo  {publisher} {Ieee Computer Soc},\ \bibinfo
  {year} {2003})\ pp.\ \bibinfo {pages} {200--209}\BibitemShut {NoStop}%
\bibitem [{\citenamefont {Grover}(1997)}]{Grover1997}%
  \BibitemOpen
  \bibfield  {author} {\bibinfo {author} {\bibfnamefont {L.~K.}\ \bibnamefont
  {Grover}},\ }\bibfield  {title} {\bibinfo {title} {Quantum mechanics helps in
  searching for a needle in a haystack},\ }\href@noop {} {\bibfield  {journal}
  {\bibinfo  {journal} {Phys. Rev. Lett.}\ }\textbf {\bibinfo {volume} {79}},\
  \bibinfo {pages} {325} (\bibinfo {year} {1997})}\BibitemShut {NoStop}%
\bibitem [{\citenamefont {Shenvi}\ \emph {et~al.}(2003)\citenamefont {Shenvi},
  \citenamefont {Kempe},\ and\ \citenamefont {Whaley}}]{Shenvi2003}%
  \BibitemOpen
  \bibfield  {author} {\bibinfo {author} {\bibfnamefont {N.}~\bibnamefont
  {Shenvi}}, \bibinfo {author} {\bibfnamefont {J.}~\bibnamefont {Kempe}},\ and\
  \bibinfo {author} {\bibfnamefont {K.~B.}\ \bibnamefont {Whaley}},\ }\bibfield
   {title} {\bibinfo {title} {Quantum random-walk search algorithm},\ }\href
  {https://doi.org/10.1103/PhysRevA.67.052307} {\bibfield  {journal} {\bibinfo
  {journal} {Phys. Rev. A}\ }\textbf {\bibinfo {volume} {67}},\ \bibinfo
  {pages} {052307} (\bibinfo {year} {2003})}\BibitemShut {NoStop}%
\bibitem [{\citenamefont {Childs}\ and\ \citenamefont
  {Goldstone}(2004{\natexlab{a}})}]{Childs2004}%
  \BibitemOpen
  \bibfield  {author} {\bibinfo {author} {\bibfnamefont {A.~M.}\ \bibnamefont
  {Childs}}\ and\ \bibinfo {author} {\bibfnamefont {J.}~\bibnamefont
  {Goldstone}},\ }\bibfield  {title} {\bibinfo {title} {Spatial search by
  quantum walk},\ }\href@noop {} {\bibfield  {journal} {\bibinfo  {journal}
  {Phys. Rev. A}\ }\textbf {\bibinfo {volume} {70}},\ \bibinfo {pages} {022314}
  (\bibinfo {year} {2004}{\natexlab{a}})}\BibitemShut {NoStop}%
\bibitem [{\citenamefont {Childs}\ and\ \citenamefont
  {Goldstone}(2004{\natexlab{b}})}]{childs_2004b}%
  \BibitemOpen
  \bibfield  {author} {\bibinfo {author} {\bibfnamefont {A.~M.}\ \bibnamefont
  {Childs}}\ and\ \bibinfo {author} {\bibfnamefont {J.}~\bibnamefont
  {Goldstone}},\ }\bibfield  {title} {\bibinfo {title} {Spatial search and the
  dirac equation},\ }\href@noop {} {\bibfield  {journal} {\bibinfo  {journal}
  {Phys. Rev. A}\ }\textbf {\bibinfo {volume} {70}},\ \bibinfo {pages} {042312}
  (\bibinfo {year} {2004}{\natexlab{b}})}\BibitemShut {NoStop}%
\bibitem [{\citenamefont {Ambainis}\ \emph {et~al.}(2005)\citenamefont
  {Ambainis}, \citenamefont {Kempe},\ and\ \citenamefont
  {Rivosh}}]{Ambainis2005}%
  \BibitemOpen
  \bibfield  {author} {\bibinfo {author} {\bibfnamefont {A.}~\bibnamefont
  {Ambainis}}, \bibinfo {author} {\bibfnamefont {J.}~\bibnamefont {Kempe}},\
  and\ \bibinfo {author} {\bibfnamefont {A.}~\bibnamefont {Rivosh}},\
  }\bibfield  {title} {\bibinfo {title} {Coins make quantum walks faster},\
  }in\ \href@noop {} {\emph {\bibinfo {booktitle} {Proc.~16th ACM-SIAM
  symposium on Discrete algorithms}}}\ (\bibinfo {organization} {Society for
  Industrial and Applied Mathematics},\ \bibinfo {year} {2005})\ pp.\ \bibinfo
  {pages} {1099--1108}\BibitemShut {NoStop}%
\bibitem [{\citenamefont {Poto{\v{c}}ek}\ \emph {et~al.}(2009)\citenamefont
  {Poto{\v{c}}ek}, \citenamefont {G{\'a}bris}, \citenamefont {Kiss},\ and\
  \citenamefont {Jex}}]{Potocek2009}%
  \BibitemOpen
  \bibfield  {author} {\bibinfo {author} {\bibfnamefont {V.}~\bibnamefont
  {Poto{\v{c}}ek}}, \bibinfo {author} {\bibfnamefont {A.}~\bibnamefont
  {G{\'a}bris}}, \bibinfo {author} {\bibfnamefont {T.}~\bibnamefont {Kiss}},\
  and\ \bibinfo {author} {\bibfnamefont {I.}~\bibnamefont {Jex}},\ }\bibfield
  {title} {\bibinfo {title} {Optimized quantum random-walk search algorithms on
  the hypercube},\ }\href@noop {} {\bibfield  {journal} {\bibinfo  {journal}
  {Phys. Rev. A}\ }\textbf {\bibinfo {volume} {79}},\ \bibinfo {pages} {012325}
  (\bibinfo {year} {2009})}\BibitemShut {NoStop}%
\bibitem [{\citenamefont {Hein}\ and\ \citenamefont
  {Tanner}(2009{\natexlab{a}})}]{hein2009:search}%
  \BibitemOpen
  \bibfield  {author} {\bibinfo {author} {\bibfnamefont {B.}~\bibnamefont
  {Hein}}\ and\ \bibinfo {author} {\bibfnamefont {G.}~\bibnamefont {Tanner}},\
  }\bibfield  {title} {\bibinfo {title} {{Quantum search algorithms on the
  hypercube}},\ }\href@noop {} {\bibfield  {journal} {\bibinfo  {journal} {{J.
  Phys. A}}\ }\textbf {\bibinfo {volume} {{42}}},\ \bibinfo {pages} {{085303}}
  (\bibinfo {year} {{2009}}{\natexlab{a}})}\BibitemShut {NoStop}%
\bibitem [{\citenamefont {Reitzner}\ \emph {et~al.}(2009)\citenamefont
  {Reitzner}, \citenamefont {Hillery}, \citenamefont {Feldman},\ and\
  \citenamefont {Bužek}}]{reitzner2009}%
  \BibitemOpen
  \bibfield  {author} {\bibinfo {author} {\bibfnamefont {D.}~\bibnamefont
  {Reitzner}}, \bibinfo {author} {\bibfnamefont {M.}~\bibnamefont {Hillery}},
  \bibinfo {author} {\bibfnamefont {E.}~\bibnamefont {Feldman}},\ and\ \bibinfo
  {author} {\bibfnamefont {V.}~\bibnamefont {Bužek}},\ }\bibfield  {title}
  {\bibinfo {title} {Quantum searches on highly symmetric graphs},\ }\href@noop
  {} {\bibfield  {journal} {\bibinfo  {journal} {Phys. Rev. A}\ }\textbf
  {\bibinfo {volume} {79}},\ \bibinfo {pages} {012323} (\bibinfo {year}
  {2009})}\BibitemShut {NoStop}%
\bibitem [{\citenamefont {Hein}\ and\ \citenamefont
  {Tanner}(2010)}]{hein_search_lattice_2010}%
  \BibitemOpen
  \bibfield  {author} {\bibinfo {author} {\bibfnamefont {B.}~\bibnamefont
  {Hein}}\ and\ \bibinfo {author} {\bibfnamefont {G.}~\bibnamefont {Tanner}},\
  }\bibfield  {title} {\bibinfo {title} {Quantum search algorithms on a regular
  lattice},\ }\href {https://doi.org/10.1103/PhysRevA.82.012326} {\bibfield
  {journal} {\bibinfo  {journal} {Phys. Rev. A}\ }\textbf {\bibinfo {volume}
  {82}},\ \bibinfo {pages} {012326} (\bibinfo {year} {2010})}\BibitemShut
  {NoStop}%
\bibitem [{\citenamefont {Wong}(2015)}]{wong2015}%
  \BibitemOpen
  \bibfield  {author} {\bibinfo {author} {\bibfnamefont {T.~G.}\ \bibnamefont
  {Wong}},\ }\bibfield  {title} {\bibinfo {title} {Grover search with
  lackadaisical quantum walks},\ }\href@noop {} {\bibfield  {journal} {\bibinfo
   {journal} {J. Phys. A}\ }\textbf {\bibinfo {volume} {48}},\ \bibinfo {pages}
  {435304} (\bibinfo {year} {2015})}\BibitemShut {NoStop}%
\bibitem [{\citenamefont {Wong}(2018)}]{wong2018}%
  \BibitemOpen
  \bibfield  {author} {\bibinfo {author} {\bibfnamefont {T.~G.}\ \bibnamefont
  {Wong}},\ }\bibfield  {title} {\bibinfo {title} {{Faster search by
  lackadaisical quantum walk}},\ }\href@noop {} {\bibfield  {journal} {\bibinfo
   {journal} {{Quantum Inf. Process.}}\ }\textbf {\bibinfo {volume} {{17}}},\
  \bibinfo {pages} {{68}} (\bibinfo {year} {{2018}})}\BibitemShut {NoStop}%
\bibitem [{\citenamefont {Rhodes}\ and\ \citenamefont
  {Wong}(2019)}]{rhodes2019}%
  \BibitemOpen
  \bibfield  {author} {\bibinfo {author} {\bibfnamefont {M.~L.}\ \bibnamefont
  {Rhodes}}\ and\ \bibinfo {author} {\bibfnamefont {T.~G.}\ \bibnamefont
  {Wong}},\ }\bibfield  {title} {\bibinfo {title} {{Search by lackadaisical
  quantum walks with nonhomogeneous weights}},\ }\href@noop {} {\bibfield
  {journal} {\bibinfo  {journal} {{Phys. Rev. A}}\ }\textbf {\bibinfo {volume}
  {{100}}},\ \bibinfo {pages} {{042303}} (\bibinfo {year}
  {{2019}})}\BibitemShut {NoStop}%
\bibitem [{\citenamefont {Rhodes}\ and\ \citenamefont
  {Wong}(2020)}]{rhodes2020}%
  \BibitemOpen
  \bibfield  {author} {\bibinfo {author} {\bibfnamefont {M.~L.}\ \bibnamefont
  {Rhodes}}\ and\ \bibinfo {author} {\bibfnamefont {T.~G.}\ \bibnamefont
  {Wong}},\ }\bibfield  {title} {\bibinfo {title} {{Search on vertex-transitive
  graphs by lackadaisical quantum walk}},\ }\href@noop {} {\bibfield  {journal}
  {\bibinfo  {journal} {{Quantum Inf. Process.}}\ }\textbf {\bibinfo {volume}
  {{19}}},\ \bibinfo {pages} {{334}} (\bibinfo {year} {{2020}})}\BibitemShut
  {NoStop}%
\bibitem [{\citenamefont {Chiang}(2020)}]{chiang2020}%
  \BibitemOpen
  \bibfield  {author} {\bibinfo {author} {\bibfnamefont {C.}~\bibnamefont
  {Chiang}},\ }\bibfield  {title} {\bibinfo {title} {Overview: recent
  development and applications of reduction and lackadaisicalness techniques
  for spatial search quantum walk in the near term},\ }\href@noop {} {\bibfield
   {journal} {\bibinfo  {journal} {Quantum Inf. Process.}\ }\textbf {\bibinfo
  {volume} {19}},\ \bibinfo {pages} {364} (\bibinfo {year} {2020})}\BibitemShut
  {NoStop}%
\bibitem [{\citenamefont {H\o{}yer}\ and\ \citenamefont
  {Yu}(2020)}]{hoyer2020}%
  \BibitemOpen
  \bibfield  {author} {\bibinfo {author} {\bibfnamefont {P.}~\bibnamefont
  {H\o{}yer}}\ and\ \bibinfo {author} {\bibfnamefont {Z.}~\bibnamefont {Yu}},\
  }\bibfield  {title} {\bibinfo {title} {{Analysis of Lackadaisical Quantum
  Walks}},\ }\href@noop {} {\bibfield  {journal} {\bibinfo  {journal} {Quant.
  Inf. Comput.}\ }\textbf {\bibinfo {volume} {20}},\ \bibinfo {pages} {14}
  (\bibinfo {year} {2020})}\BibitemShut {NoStop}%
\bibitem [{\citenamefont {Chakraborty}\ \emph {et~al.}(2016)\citenamefont
  {Chakraborty}, \citenamefont {Novo}, \citenamefont {Ambainis},\ and\
  \citenamefont {Omar}}]{chakraborty2016}%
  \BibitemOpen
  \bibfield  {author} {\bibinfo {author} {\bibfnamefont {S.}~\bibnamefont
  {Chakraborty}}, \bibinfo {author} {\bibfnamefont {L.}~\bibnamefont {Novo}},
  \bibinfo {author} {\bibfnamefont {A.}~\bibnamefont {Ambainis}},\ and\
  \bibinfo {author} {\bibfnamefont {Y.}~\bibnamefont {Omar}},\ }\bibfield
  {title} {\bibinfo {title} {Spatial search by quantum walk is optimal for
  almost all graphs},\ }\href@noop {} {\bibfield  {journal} {\bibinfo
  {journal} {Phys. Rev. Lett.}\ }\textbf {\bibinfo {volume} {116}},\ \bibinfo
  {pages} {100501} (\bibinfo {year} {2016})}\BibitemShut {NoStop}%
\bibitem [{\citenamefont {Ambainis}\ \emph {et~al.}(2020)\citenamefont
  {Ambainis}, \citenamefont {Gilyen}, \citenamefont {Jeffery},\ and\
  \citenamefont {Kokainis}}]{ambainis_quadratic_2020}%
  \BibitemOpen
  \bibfield  {author} {\bibinfo {author} {\bibfnamefont {A.}~\bibnamefont
  {Ambainis}}, \bibinfo {author} {\bibfnamefont {A.}~\bibnamefont {Gilyen}},
  \bibinfo {author} {\bibfnamefont {S.}~\bibnamefont {Jeffery}},\ and\ \bibinfo
  {author} {\bibfnamefont {M.}~\bibnamefont {Kokainis}},\ }\bibfield  {title}
  {\bibinfo {title} {Quadratic {Speedup} for {Finding} {Marked} {Vertices} by
  {Quantum} {Walks}},\ }in\ \href@noop {} {\emph {\bibinfo {booktitle}
  {Proceedings of the 52nd {Annual} {Acm} {Sigact} {Symposium} on {Theory} of
  {Computing} (stoc '20)}}}\ (\bibinfo  {publisher} {Assoc Computing
  Machinery},\ \bibinfo {year} {2020})\ pp.\ \bibinfo {pages}
  {412--424}\BibitemShut {NoStop}%
\bibitem [{\citenamefont {Apers}\ \emph {et~al.}(2022)\citenamefont {Apers},
  \citenamefont {Chakraborty}, \citenamefont {Novo},\ and\ \citenamefont
  {Roland}}]{apers_quadratic_2022}%
  \BibitemOpen
  \bibfield  {author} {\bibinfo {author} {\bibfnamefont {S.}~\bibnamefont
  {Apers}}, \bibinfo {author} {\bibfnamefont {S.}~\bibnamefont {Chakraborty}},
  \bibinfo {author} {\bibfnamefont {L.}~\bibnamefont {Novo}},\ and\ \bibinfo
  {author} {\bibfnamefont {J.}~\bibnamefont {Roland}},\ }\bibfield  {title}
  {\bibinfo {title} {Quadratic {Speedup} for {Spatial} {Search} by
  {Continuous}-{Time} {Quantum} {Walk}},\ }\href@noop {} {\bibfield  {journal}
  {\bibinfo  {journal} {Phys. Rev. Lett.}\ }\textbf {\bibinfo {volume} {129}},\
  \bibinfo {pages} {160502} (\bibinfo {year} {2022})}\BibitemShut {NoStop}%
\bibitem [{\citenamefont {Bose}(2003)}]{bose2003}%
  \BibitemOpen
  \bibfield  {author} {\bibinfo {author} {\bibfnamefont {S.}~\bibnamefont
  {Bose}},\ }\bibfield  {title} {\bibinfo {title} {Quantum communication
  through an unmodulated spin chain},\ }\href@noop {} {\bibfield  {journal}
  {\bibinfo  {journal} {Phys. Rev. Lett.}\ }\textbf {\bibinfo {volume} {91}},\
  \bibinfo {pages} {207901} (\bibinfo {year} {2003})}\BibitemShut {NoStop}%
\bibitem [{\citenamefont {Gisin}\ and\ \citenamefont
  {Thew}(2007)}]{Gisin_QCom_2007}%
  \BibitemOpen
  \bibfield  {author} {\bibinfo {author} {\bibfnamefont {N.}~\bibnamefont
  {Gisin}}\ and\ \bibinfo {author} {\bibfnamefont {R.}~\bibnamefont {Thew}},\
  }\bibfield  {title} {\bibinfo {title} {Quantum communication},\ }\href
  {https://doi.org/10.1038/nphoton.2007.22} {\bibfield  {journal} {\bibinfo
  {journal} {Nature Photonics}\ }\textbf {\bibinfo {volume} {1}},\ \bibinfo
  {pages} {165} (\bibinfo {year} {2007})}\BibitemShut {NoStop}%
\bibitem [{\citenamefont {Buhrman}\ and\ \citenamefont
  {Röhrig}(2003)}]{Buhrman_DQC_2003}%
  \BibitemOpen
  \bibfield  {author} {\bibinfo {author} {\bibfnamefont {H.}~\bibnamefont
  {Buhrman}}\ and\ \bibinfo {author} {\bibfnamefont {H.}~\bibnamefont
  {Röhrig}},\ }\bibfield  {title} {\bibinfo {title} {Distributed quantum
  computing},\ }in\ \href@noop {} {\emph {\bibinfo {booktitle} {Proceedings of
  28th International Symposium on Mathematical Foundations of Computer
  Science}}},\ \bibinfo {series} {Lecture Notes in Computer Science}, Vol.\
  \bibinfo {volume} {2747},\ \bibinfo {editor} {edited by\ \bibinfo {editor}
  {\bibfnamefont {B.}~\bibnamefont {Rovan}}\ and\ \bibinfo {editor}
  {\bibfnamefont {P.}~\bibnamefont {Vojtas}}}\ (\bibinfo {year} {2003})\ pp.\
  \bibinfo {pages} {1--20}\BibitemShut {NoStop}%
\bibitem [{\citenamefont {Cuomo}\ \emph {et~al.}(2020)\citenamefont {Cuomo},
  \citenamefont {Caleffi},\ and\ \citenamefont
  {Cacciapuoti}}]{Cuomo_distributed_QC_2020}%
  \BibitemOpen
  \bibfield  {author} {\bibinfo {author} {\bibfnamefont {D.}~\bibnamefont
  {Cuomo}}, \bibinfo {author} {\bibfnamefont {M.}~\bibnamefont {Caleffi}},\
  and\ \bibinfo {author} {\bibfnamefont {A.~S.}\ \bibnamefont {Cacciapuoti}},\
  }\bibfield  {title} {\bibinfo {title} {Towards a distributed quantum
  computing ecosystem},\ }\href {https://doi.org/10.1049/iet-qtc.2020.0002}
  {\bibfield  {journal} {\bibinfo  {journal} {IET Quantum Commun.}\ }\textbf
  {\bibinfo {volume} {1}},\ \bibinfo {pages} {3} (\bibinfo {year}
  {2020})}\BibitemShut {NoStop}%
\bibitem [{\citenamefont {Kimble}(2008)}]{Kimble_QI_2008}%
  \BibitemOpen
  \bibfield  {author} {\bibinfo {author} {\bibfnamefont {H.~J.}\ \bibnamefont
  {Kimble}},\ }\bibfield  {title} {\bibinfo {title} {The quantum internet},\
  }\href {https://doi.org/10.1038/nature07127} {\bibfield  {journal} {\bibinfo
  {journal} {Nature}\ }\textbf {\bibinfo {volume} {453}},\ \bibinfo {pages}
  {1023} (\bibinfo {year} {2008})}\BibitemShut {NoStop}%
\bibitem [{\citenamefont {Caleffi}\ \emph {et~al.}(2018)\citenamefont
  {Caleffi}, \citenamefont {Cacciapuoti},\ and\ \citenamefont
  {Bianchi}}]{Caleffi_QI_communication_2018}%
  \BibitemOpen
  \bibfield  {author} {\bibinfo {author} {\bibfnamefont {M.}~\bibnamefont
  {Caleffi}}, \bibinfo {author} {\bibfnamefont {A.~S.}\ \bibnamefont
  {Cacciapuoti}},\ and\ \bibinfo {author} {\bibfnamefont {G.}~\bibnamefont
  {Bianchi}},\ }\bibfield  {title} {\bibinfo {title} {Quantum internet: from
  communication to distributed computing!},\ }in\ \href
  {https://doi.org/10.1145/3233188.3233224} {\emph {\bibinfo {booktitle} {ACM
  NANOCOM 2018: 5th ACM International Conference on Nanoscale Computing and
  Communication}}}\ (\bibinfo {organization} {ACM},\ \bibinfo {year}
  {2018})\BibitemShut {NoStop}%
\bibitem [{\citenamefont {Rohde}(2021)}]{Rohde_quantum_internet_2021}%
  \BibitemOpen
  \bibfield  {author} {\bibinfo {author} {\bibfnamefont {P.~P.}\ \bibnamefont
  {Rohde}},\ }\href@noop {} {\emph {\bibinfo {title} {The Quantum Internet: The
  Second Quantum Revolution}}}\ (\bibinfo  {publisher} {Cambridge University
  Press},\ \bibinfo {year} {2021})\BibitemShut {NoStop}%
\bibitem [{\citenamefont {Christandl}\ \emph {et~al.}(2004)\citenamefont
  {Christandl}, \citenamefont {Datta}, \citenamefont {Ekert},\ and\
  \citenamefont {Landahl}}]{christandl_perfect_2004}%
  \BibitemOpen
  \bibfield  {author} {\bibinfo {author} {\bibfnamefont {M.}~\bibnamefont
  {Christandl}}, \bibinfo {author} {\bibfnamefont {N.}~\bibnamefont {Datta}},
  \bibinfo {author} {\bibfnamefont {A.}~\bibnamefont {Ekert}},\ and\ \bibinfo
  {author} {\bibfnamefont {A.~J.}\ \bibnamefont {Landahl}},\ }\bibfield
  {title} {\bibinfo {title} {Perfect {State} {Transfer} in {Quantum} {Spin}
  {Networks}},\ }\href@noop {} {\bibfield  {journal} {\bibinfo  {journal}
  {Phys. Rev. Lett.}\ }\textbf {\bibinfo {volume} {92}},\ \bibinfo {pages}
  {187902} (\bibinfo {year} {2004})}\BibitemShut {NoStop}%
\bibitem [{\citenamefont {Christandl}\ \emph {et~al.}(2005)\citenamefont
  {Christandl}, \citenamefont {Datta}, \citenamefont {Dorlas}, \citenamefont
  {Ekert}, \citenamefont {Kay},\ and\ \citenamefont
  {Landahl}}]{christandl_perfect_2005}%
  \BibitemOpen
  \bibfield  {author} {\bibinfo {author} {\bibfnamefont {M.}~\bibnamefont
  {Christandl}}, \bibinfo {author} {\bibfnamefont {N.}~\bibnamefont {Datta}},
  \bibinfo {author} {\bibfnamefont {T.~C.}\ \bibnamefont {Dorlas}}, \bibinfo
  {author} {\bibfnamefont {A.}~\bibnamefont {Ekert}}, \bibinfo {author}
  {\bibfnamefont {A.}~\bibnamefont {Kay}},\ and\ \bibinfo {author}
  {\bibfnamefont {A.~J.}\ \bibnamefont {Landahl}},\ }\bibfield  {title}
  {\bibinfo {title} {Perfect transfer of arbitrary states in quantum spin
  networks},\ }\href@noop {} {\bibfield  {journal} {\bibinfo  {journal} {Phys.
  Rev. A}\ }\textbf {\bibinfo {volume} {71}},\ \bibinfo {pages} {032312}
  (\bibinfo {year} {2005})}\BibitemShut {NoStop}%
\bibitem [{\citenamefont {Plenio}\ and\ \citenamefont
  {Semião}(2005)}]{plenio_high_2005}%
  \BibitemOpen
  \bibfield  {author} {\bibinfo {author} {\bibfnamefont {M.~B.}\ \bibnamefont
  {Plenio}}\ and\ \bibinfo {author} {\bibfnamefont {F.~L.}\ \bibnamefont
  {Semião}},\ }\bibfield  {title} {\bibinfo {title} {High efficiency transfer
  of quantum information and multiparticle entanglement generation in
  translation-invariant quantum chains},\ }\href@noop {} {\bibfield  {journal}
  {\bibinfo  {journal} {New J. Phys.}\ }\textbf {\bibinfo {volume} {7}},\
  \bibinfo {pages} {73} (\bibinfo {year} {2005})}\BibitemShut {NoStop}%
\bibitem [{\citenamefont {Bose}(2007)}]{bose_2007}%
  \BibitemOpen
  \bibfield  {author} {\bibinfo {author} {\bibfnamefont {S.}~\bibnamefont
  {Bose}},\ }\bibfield  {title} {\bibinfo {title} {Quantum communication
  through spin chain dynamics: an introductory overview},\ }\href@noop {}
  {\bibfield  {journal} {\bibinfo  {journal} {Contemp. Phys.}\ }\textbf
  {\bibinfo {volume} {48}},\ \bibinfo {pages} {13} (\bibinfo {year}
  {2007})}\BibitemShut {NoStop}%
\bibitem [{\citenamefont {Kostak}\ \emph {et~al.}(2007)\citenamefont {Kostak},
  \citenamefont {Nikolopoulos},\ and\ \citenamefont
  {Jex}}]{kostak_perfect_2007}%
  \BibitemOpen
  \bibfield  {author} {\bibinfo {author} {\bibfnamefont {V.}~\bibnamefont
  {Kostak}}, \bibinfo {author} {\bibfnamefont {G.~M.}\ \bibnamefont
  {Nikolopoulos}},\ and\ \bibinfo {author} {\bibfnamefont {I.}~\bibnamefont
  {Jex}},\ }\bibfield  {title} {\bibinfo {title} {Perfect state transfer in
  networks of arbitrary topology and coupling configuration},\ }\href@noop {}
  {\bibfield  {journal} {\bibinfo  {journal} {Phys. Rev. A}\ }\textbf {\bibinfo
  {volume} {75}},\ \bibinfo {pages} {042319} (\bibinfo {year}
  {2007})}\BibitemShut {NoStop}%
\bibitem [{\citenamefont {Gualdi}\ \emph {et~al.}(2008)\citenamefont {Gualdi},
  \citenamefont {Kostak}, \citenamefont {Marzoli},\ and\ \citenamefont
  {Tombesi}}]{gualdi_perfect_2008}%
  \BibitemOpen
  \bibfield  {author} {\bibinfo {author} {\bibfnamefont {G.}~\bibnamefont
  {Gualdi}}, \bibinfo {author} {\bibfnamefont {V.}~\bibnamefont {Kostak}},
  \bibinfo {author} {\bibfnamefont {I.}~\bibnamefont {Marzoli}},\ and\ \bibinfo
  {author} {\bibfnamefont {P.}~\bibnamefont {Tombesi}},\ }\bibfield  {title}
  {\bibinfo {title} {Perfect state transfer in long-range interacting spin
  chains},\ }\href@noop {} {\bibfield  {journal} {\bibinfo  {journal} {Phys.
  Rev. A}\ }\textbf {\bibinfo {volume} {78}},\ \bibinfo {pages} {022325}
  (\bibinfo {year} {2008})}\BibitemShut {NoStop}%
\bibitem [{\citenamefont {Kay}(2010)}]{kay_perfect_2010}%
  \BibitemOpen
  \bibfield  {author} {\bibinfo {author} {\bibfnamefont {A.}~\bibnamefont
  {Kay}},\ }\bibfield  {title} {\bibinfo {title} {{Perfect}, {efficient},
  {state} {transfer} {and} {its} {application} {as} {a} {constructive}
  {tool}},\ }\href@noop {} {\bibfield  {journal} {\bibinfo  {journal} {Int. J.
  Quantum Inf.}\ }\textbf {\bibinfo {volume} {8}},\ \bibinfo {pages} {641}
  (\bibinfo {year} {2010})}\BibitemShut {NoStop}%
\bibitem [{\citenamefont {Kendon}\ and\ \citenamefont
  {Tamon}(2011)}]{kendon2011}%
  \BibitemOpen
  \bibfield  {author} {\bibinfo {author} {\bibfnamefont {V.~M.}\ \bibnamefont
  {Kendon}}\ and\ \bibinfo {author} {\bibfnamefont {C.}~\bibnamefont {Tamon}},\
  }\bibfield  {title} {\bibinfo {title} {{Perfect state transfer in quantum
  walks on graphs}},\ }\href@noop {} {\bibfield  {journal} {\bibinfo  {journal}
  {{J. Comput. Theor. Nanosci.}}\ }\textbf {\bibinfo {volume} {{8}}},\ \bibinfo
  {pages} {{422}} (\bibinfo {year} {{2011}})}\BibitemShut {NoStop}%
\bibitem [{\citenamefont {Godsil}(2012)}]{godsil_state_2012}%
  \BibitemOpen
  \bibfield  {author} {\bibinfo {author} {\bibfnamefont {C.}~\bibnamefont
  {Godsil}},\ }\bibfield  {title} {\bibinfo {title} {State transfer on
  graphs},\ }\href@noop {} {\bibfield  {journal} {\bibinfo  {journal} {Discrete
  Math.}\ }\textbf {\bibinfo {volume} {312}},\ \bibinfo {pages} {129} (\bibinfo
  {year} {2012})}\BibitemShut {NoStop}%
\bibitem [{\citenamefont {Nikolopoulos}\ \emph {et~al.}(2012)\citenamefont
  {Nikolopoulos}, \citenamefont {Hoskovec},\ and\ \citenamefont
  {Jex}}]{nikolopoulos_analysis_2012}%
  \BibitemOpen
  \bibfield  {author} {\bibinfo {author} {\bibfnamefont {G.~M.}\ \bibnamefont
  {Nikolopoulos}}, \bibinfo {author} {\bibfnamefont {A.}~\bibnamefont
  {Hoskovec}},\ and\ \bibinfo {author} {\bibfnamefont {I.}~\bibnamefont
  {Jex}},\ }\bibfield  {title} {\bibinfo {title} {Analysis and minimization of
  bending losses in discrete quantum networks},\ }\href@noop {} {\bibfield
  {journal} {\bibinfo  {journal} {Phys. Rev. A}\ }\textbf {\bibinfo {volume}
  {85}},\ \bibinfo {pages} {062319} (\bibinfo {year} {2012})}\BibitemShut
  {NoStop}%
\bibitem [{\citenamefont {Hoskovec}\ \emph {et~al.}(2014)\citenamefont
  {Hoskovec}, \citenamefont {Frydrych}, \citenamefont {Jex},\ and\
  \citenamefont {Alber}}]{hoskovec_decoupling_2014}%
  \BibitemOpen
  \bibfield  {author} {\bibinfo {author} {\bibfnamefont {A.}~\bibnamefont
  {Hoskovec}}, \bibinfo {author} {\bibfnamefont {H.}~\bibnamefont {Frydrych}},
  \bibinfo {author} {\bibfnamefont {I.}~\bibnamefont {Jex}},\ and\ \bibinfo
  {author} {\bibfnamefont {G.}~\bibnamefont {Alber}},\ }\bibfield  {title}
  {\bibinfo {title} {Decoupling {Bent} {Quantum} {Networks}},\ }in\ \href@noop
  {} {\emph {\bibinfo {booktitle} {2014 {International} {Symposium} on
  {Information} {Theory} and {Its} {Applications} (isita)}}}\ (\bibinfo
  {publisher} {IEEE},\ \bibinfo {address} {New York},\ \bibinfo {year} {2014})\
  pp.\ \bibinfo {pages} {172--175}\BibitemShut {NoStop}%
\bibitem [{\citenamefont {Frydrych}\ \emph {et~al.}(2015)\citenamefont
  {Frydrych}, \citenamefont {Hoskovec}, \citenamefont {Jex},\ and\
  \citenamefont {Alber}}]{frydrych_selective_2015}%
  \BibitemOpen
  \bibfield  {author} {\bibinfo {author} {\bibfnamefont {H.}~\bibnamefont
  {Frydrych}}, \bibinfo {author} {\bibfnamefont {A.}~\bibnamefont {Hoskovec}},
  \bibinfo {author} {\bibfnamefont {I.}~\bibnamefont {Jex}},\ and\ \bibinfo
  {author} {\bibfnamefont {G.}~\bibnamefont {Alber}},\ }\bibfield  {title}
  {\bibinfo {title} {Selective dynamical decoupling for quantum state
  transfer},\ }\href@noop {} {\bibfield  {journal} {\bibinfo  {journal} {J.
  Phys. B - At. Mol. Opt. Phys.}\ }\textbf {\bibinfo {volume} {48}},\ \bibinfo
  {pages} {025501} (\bibinfo {year} {2015})}\BibitemShut {NoStop}%
\bibitem [{\citenamefont {Coutinho}\ \emph {et~al.}(2015)\citenamefont
  {Coutinho}, \citenamefont {Godsil}, \citenamefont {Guo},\ and\ \citenamefont
  {Vanhove}}]{coutinho_perfect_2015}%
  \BibitemOpen
  \bibfield  {author} {\bibinfo {author} {\bibfnamefont {G.}~\bibnamefont
  {Coutinho}}, \bibinfo {author} {\bibfnamefont {C.}~\bibnamefont {Godsil}},
  \bibinfo {author} {\bibfnamefont {K.}~\bibnamefont {Guo}},\ and\ \bibinfo
  {author} {\bibfnamefont {F.}~\bibnamefont {Vanhove}},\ }\bibfield  {title}
  {\bibinfo {title} {Perfect state transfer on distance-regular graphs and
  association schemes},\ }\href@noop {} {\bibfield  {journal} {\bibinfo
  {journal} {Linear Algebra Appl.}\ }\textbf {\bibinfo {volume} {478}},\
  \bibinfo {pages} {108} (\bibinfo {year} {2015})}\BibitemShut {NoStop}%
\bibitem [{\citenamefont {Li~et al.}(2018)}]{li_qubit_chain_2018}%
  \BibitemOpen
  \bibfield  {author} {\bibinfo {author} {\bibfnamefont {X.}~\bibnamefont
  {Li~et al.}},\ }\bibfield  {title} {\bibinfo {title} {Perfect quantum state
  transfer in a superconducting qubit chain with parametrically tunable
  couplings},\ }\href@noop {} {\bibfield  {journal} {\bibinfo  {journal} {Phys.
  Rev. Appl.}\ }\textbf {\bibinfo {volume} {10}},\ \bibinfo {pages} {054009}
  (\bibinfo {year} {2018})}\BibitemShut {NoStop}%
\bibitem [{\citenamefont {Godsil}\ \emph {et~al.}(2020)\citenamefont {Godsil},
  \citenamefont {Guo}, \citenamefont {Kempton}, \citenamefont {Lippner},\ and\
  \citenamefont {Münch}}]{godsil_state_2020}%
  \BibitemOpen
  \bibfield  {author} {\bibinfo {author} {\bibfnamefont {C.}~\bibnamefont
  {Godsil}}, \bibinfo {author} {\bibfnamefont {K.}~\bibnamefont {Guo}},
  \bibinfo {author} {\bibfnamefont {M.}~\bibnamefont {Kempton}}, \bibinfo
  {author} {\bibfnamefont {G.}~\bibnamefont {Lippner}},\ and\ \bibinfo {author}
  {\bibfnamefont {F.}~\bibnamefont {Münch}},\ }\bibfield  {title} {\bibinfo
  {title} {State transfer in strongly regular graphs with an edge
  perturbation},\ }\href@noop {} {\bibfield  {journal} {\bibinfo  {journal} {J.
  Comb. Theory, Series A}\ }\textbf {\bibinfo {volume} {172}},\ \bibinfo
  {pages} {105181} (\bibinfo {year} {2020})}\BibitemShut {NoStop}%
\bibitem [{\citenamefont {Chen}\ and\ \citenamefont
  {Godsil}(2020)}]{chen_pair_2020}%
  \BibitemOpen
  \bibfield  {author} {\bibinfo {author} {\bibfnamefont {Q.}~\bibnamefont
  {Chen}}\ and\ \bibinfo {author} {\bibfnamefont {C.}~\bibnamefont {Godsil}},\
  }\bibfield  {title} {\bibinfo {title} {Pair state transfer},\ }\href@noop {}
  {\bibfield  {journal} {\bibinfo  {journal} {Quantum Inf. Process.}\ }\textbf
  {\bibinfo {volume} {19}},\ \bibinfo {pages} {321} (\bibinfo {year}
  {2020})}\BibitemShut {NoStop}%
\bibitem [{\citenamefont {Cao}(2021)}]{cao_perfect_2021}%
  \BibitemOpen
  \bibfield  {author} {\bibinfo {author} {\bibfnamefont {X.}~\bibnamefont
  {Cao}},\ }\bibfield  {title} {\bibinfo {title} {Perfect edge state transfer
  on cubelike graphs},\ }\href {https://doi.org/10.1007/s11128-021-03223-9}
  {\bibfield  {journal} {\bibinfo  {journal} {Quantum Inform. Process.}\
  }\textbf {\bibinfo {volume} {20}},\ \bibinfo {pages} {285} (\bibinfo {year}
  {2021})}\BibitemShut {NoStop}%
\bibitem [{\citenamefont {Hoskovec}\ and\ \citenamefont
  {Jex}(2022)}]{hoskovec_dynamical_2022}%
  \BibitemOpen
  \bibfield  {author} {\bibinfo {author} {\bibfnamefont {A.}~\bibnamefont
  {Hoskovec}}\ and\ \bibinfo {author} {\bibfnamefont {I.}~\bibnamefont {Jex}},\
  }\bibfield  {title} {\bibinfo {title} {Dynamical decoupling and {NNN}
  discrete quantum networks},\ }\href@noop {} {\bibfield  {journal} {\bibinfo
  {journal} {Int. J. Quantum Inf.}\ }\textbf {\bibinfo {volume} {20}},\
  \bibinfo {pages} {2250009} (\bibinfo {year} {2022})}\BibitemShut {NoStop}%
\bibitem [{\citenamefont {Cao}\ and\ \citenamefont
  {Wan}(2022)}]{cao_perfect_2022}%
  \BibitemOpen
  \bibfield  {author} {\bibinfo {author} {\bibfnamefont {X.}~\bibnamefont
  {Cao}}\ and\ \bibinfo {author} {\bibfnamefont {J.}~\bibnamefont {Wan}},\
  }\bibfield  {title} {\bibinfo {title} {Perfect edge state transfer on abelian
  {Cayley} graphs},\ }\href {https://doi.org/10.1016/j.laa.2022.08.003}
  {\bibfield  {journal} {\bibinfo  {journal} {Linear Algebra Appl.}\ }\textbf
  {\bibinfo {volume} {653}},\ \bibinfo {pages} {44} (\bibinfo {year}
  {2022})}\BibitemShut {NoStop}%
\bibitem [{\citenamefont {Arezoomand}\ \emph {et~al.}(2022)\citenamefont
  {Arezoomand}, \citenamefont {Shafiei},\ and\ \citenamefont
  {Ghorbani}}]{arezoomand_perfect_2022}%
  \BibitemOpen
  \bibfield  {author} {\bibinfo {author} {\bibfnamefont {M.}~\bibnamefont
  {Arezoomand}}, \bibinfo {author} {\bibfnamefont {F.}~\bibnamefont
  {Shafiei}},\ and\ \bibinfo {author} {\bibfnamefont {M.}~\bibnamefont
  {Ghorbani}},\ }\bibfield  {title} {\bibinfo {title} {Perfect state transfer
  on {Cayley} graphs over the dicyclic group},\ }\href
  {https://doi.org/10.1016/j.laa.2021.12.019} {\bibfield  {journal} {\bibinfo
  {journal} {Linear Algebra Appl.}\ }\textbf {\bibinfo {volume} {639}},\
  \bibinfo {pages} {116} (\bibinfo {year} {2022})}\BibitemShut {NoStop}%
\bibitem [{\citenamefont {Wang}\ and\ \citenamefont
  {Cao}(2022)}]{wang_perfect_2022}%
  \BibitemOpen
  \bibfield  {author} {\bibinfo {author} {\bibfnamefont {D.}~\bibnamefont
  {Wang}}\ and\ \bibinfo {author} {\bibfnamefont {X.}~\bibnamefont {Cao}},\
  }\bibfield  {title} {\bibinfo {title} {Perfect quantum state transfer on
  {Cayley} graphs over dicyclic groups},\ }\href
  {https://doi.org/10.1080/03081087.2022.2158163} {\bibfield  {journal}
  {\bibinfo  {journal} {Linear Multilinear Algebra}\ }\textbf {\bibinfo
  {volume} {72}},\ \bibinfo {pages} {76} (\bibinfo {year} {2022})}\BibitemShut
  {NoStop}%
\bibitem [{\citenamefont {Wang}\ and\ \citenamefont
  {Feng}(2023)}]{wang_perfect_2023}%
  \BibitemOpen
  \bibfield  {author} {\bibinfo {author} {\bibfnamefont {S.}~\bibnamefont
  {Wang}}\ and\ \bibinfo {author} {\bibfnamefont {T.}~\bibnamefont {Feng}},\
  }\bibfield  {title} {\bibinfo {title} {Perfect state transfer on bi-{Cayley}
  graphs over abelian groups},\ }\href
  {https://doi.org/10.1016/j.disc.2023.113362} {\bibfield  {journal} {\bibinfo
  {journal} {Disc. Math.}\ }\textbf {\bibinfo {volume} {346}},\ \bibinfo
  {pages} {113362} (\bibinfo {year} {2023})}\BibitemShut {NoStop}%
\bibitem [{\citenamefont {Wang}(2023)}]{wang_abstract_2023}%
  \BibitemOpen
  \bibfield  {author} {\bibinfo {author} {\bibfnamefont {C.}~\bibnamefont
  {Wang}},\ }\bibfield  {title} {\bibinfo {title} {Abstract model of
  continuous-time quantum walk based on {Bernoulli} functionals and perfect
  state transfer},\ }\href {https://doi.org/10.1142/S0219749923500156}
  {\bibfield  {journal} {\bibinfo  {journal} {Int. J. Quantum Inform.}\
  }\textbf {\bibinfo {volume} {21}},\ \bibinfo {pages} {2350015} (\bibinfo
  {year} {2023})}\BibitemShut {NoStop}%
\bibitem [{\citenamefont {Arezoomand}(2023)}]{arezoomand_perfect_2023}%
  \BibitemOpen
  \bibfield  {author} {\bibinfo {author} {\bibfnamefont {M.}~\bibnamefont
  {Arezoomand}},\ }\bibfield  {title} {\bibinfo {title} {Perfect state transfer
  on semi-{Cayley} graphs over abelian groups},\ }\href
  {https://doi.org/10.1080/03081087.2022.2101602} {\bibfield  {journal}
  {\bibinfo  {journal} {Linear Multilinear Algebra}\ }\textbf {\bibinfo
  {volume} {71}},\ \bibinfo {pages} {2337} (\bibinfo {year}
  {2023})}\BibitemShut {NoStop}%
\bibitem [{\citenamefont {Kurzynski}\ and\ \citenamefont
  {Wojcik}(2011)}]{kurzynski2011}%
  \BibitemOpen
  \bibfield  {author} {\bibinfo {author} {\bibfnamefont {P.}~\bibnamefont
  {Kurzynski}}\ and\ \bibinfo {author} {\bibfnamefont {A.}~\bibnamefont
  {Wojcik}},\ }\bibfield  {title} {\bibinfo {title} {{Discrete-time quantum
  walk approach to state transfer}},\ }\href@noop {} {\bibfield  {journal}
  {\bibinfo  {journal} {{Phys. Rev. A}}\ }\textbf {\bibinfo {volume} {{83}}},\
  \bibinfo {pages} {{062315}} (\bibinfo {year} {{2011}})}\BibitemShut {NoStop}%
\bibitem [{\citenamefont {Yalcinkaya}\ and\ \citenamefont
  {Gedik}(2015)}]{yalcinkaya2015}%
  \BibitemOpen
  \bibfield  {author} {\bibinfo {author} {\bibfnamefont {I.}~\bibnamefont
  {Yalcinkaya}}\ and\ \bibinfo {author} {\bibfnamefont {Z.}~\bibnamefont
  {Gedik}},\ }\bibfield  {title} {\bibinfo {title} {{Qubit state transfer via
  discrete-time quantum walks}},\ }\href@noop {} {\bibfield  {journal}
  {\bibinfo  {journal} {{J. Phys. A}}\ }\textbf {\bibinfo {volume} {{48}}},\
  \bibinfo {pages} {{225302}} (\bibinfo {year} {{2015}})}\BibitemShut {NoStop}%
\bibitem [{\citenamefont {Shang}\ \emph {et~al.}(2018)\citenamefont {Shang},
  \citenamefont {Wang}, \citenamefont {Li},\ and\ \citenamefont
  {Lu}}]{shang2018}%
  \BibitemOpen
  \bibfield  {author} {\bibinfo {author} {\bibfnamefont {Y.}~\bibnamefont
  {Shang}}, \bibinfo {author} {\bibfnamefont {Y.}~\bibnamefont {Wang}},
  \bibinfo {author} {\bibfnamefont {M.}~\bibnamefont {Li}},\ and\ \bibinfo
  {author} {\bibfnamefont {R.~Q.}\ \bibnamefont {Lu}},\ }\bibfield  {title}
  {\bibinfo {title} {Quantum communication protocols by quantum walks with two
  coins},\ }\href@noop {} {\bibfield  {journal} {\bibinfo  {journal} {EPL}\
  }\textbf {\bibinfo {volume} {124}},\ \bibinfo {pages} {60009} (\bibinfo
  {year} {2018})}\BibitemShut {NoStop}%
\bibitem [{\citenamefont {Chen}\ \emph {et~al.}(2019)\citenamefont {Chen},
  \citenamefont {Wang}, \citenamefont {Xu},\ and\ \citenamefont
  {Yang}}]{chen2019}%
  \BibitemOpen
  \bibfield  {author} {\bibinfo {author} {\bibfnamefont {X.~B.}\ \bibnamefont
  {Chen}}, \bibinfo {author} {\bibfnamefont {Y.~L.}\ \bibnamefont {Wang}},
  \bibinfo {author} {\bibfnamefont {G.}~\bibnamefont {Xu}},\ and\ \bibinfo
  {author} {\bibfnamefont {Y.~X.}\ \bibnamefont {Yang}},\ }\bibfield  {title}
  {\bibinfo {title} {Quantum network communication with a novel discrete-time
  quantum walk},\ }\href@noop {} {\bibfield  {journal} {\bibinfo  {journal}
  {IEEE Access}\ }\textbf {\bibinfo {volume} {7}},\ \bibinfo {pages} {13634}
  (\bibinfo {year} {2019})}\BibitemShut {NoStop}%
\bibitem [{\citenamefont {Zhan}(2021)}]{zhan_quantum_2021}%
  \BibitemOpen
  \bibfield  {author} {\bibinfo {author} {\bibfnamefont {H.}~\bibnamefont
  {Zhan}},\ }\bibfield  {title} {\bibinfo {title} {Quantum walks on
  embeddings},\ }\href@noop {} {\bibfield  {journal} {\bibinfo  {journal} {J.
  Algebr. Comb.}\ }\textbf {\bibinfo {volume} {53}},\ \bibinfo {pages} {1187}
  (\bibinfo {year} {2021})}\BibitemShut {NoStop}%
\bibitem [{\citenamefont {Kubota}\ and\ \citenamefont
  {Segawa}(2022)}]{kubota_perfect_2022}%
  \BibitemOpen
  \bibfield  {author} {\bibinfo {author} {\bibfnamefont {S.}~\bibnamefont
  {Kubota}}\ and\ \bibinfo {author} {\bibfnamefont {E.}~\bibnamefont
  {Segawa}},\ }\bibfield  {title} {\bibinfo {title} {Perfect state transfer in
  {Grover} walks between states associated to vertices of a graph},\
  }\href@noop {} {\bibfield  {journal} {\bibinfo  {journal} {Linear Algebra
  Appl.}\ }\textbf {\bibinfo {volume} {646}},\ \bibinfo {pages} {238} (\bibinfo
  {year} {2022})}\BibitemShut {NoStop}%
\bibitem [{\citenamefont {Guo}\ and\ \citenamefont {Schmeits}()}]{guo:2022}%
  \BibitemOpen
  \bibfield  {author} {\bibinfo {author} {\bibfnamefont {K.}~\bibnamefont
  {Guo}}\ and\ \bibinfo {author} {\bibfnamefont {V.}~\bibnamefont {Schmeits}},\
  }\bibfield  {title} {\bibinfo {title} {Perfect state transfer in quantum
  walks on orientable maps},\ }\href@noop {} {\bibinfo  {journal}
  {arXiv:2211.12841}\ }\BibitemShut {NoStop}%
\bibitem [{\citenamefont {Chan}\ and\ \citenamefont {Zhan}(2023)}]{chan:2023}%
  \BibitemOpen
\bibfield  {journal} {  }\bibfield  {author} {\bibinfo {author} {\bibfnamefont
  {A.}~\bibnamefont {Chan}}\ and\ \bibinfo {author} {\bibfnamefont
  {H.}~\bibnamefont {Zhan}},\ }\bibfield  {title} {\bibinfo {title} {Pretty
  good state transfer in discrete-time quantum walks},\ }\href@noop {}
  {\bibfield  {journal} {\bibinfo  {journal} {J. Phys. A: Math. Theor.}\
  }\textbf {\bibinfo {volume} {56}},\ \bibinfo {pages} {165305} (\bibinfo
  {year} {2023})}\BibitemShut {NoStop}%
\bibitem [{\citenamefont {Hein}\ and\ \citenamefont
  {Tanner}(2009{\natexlab{b}})}]{hein2009}%
  \BibitemOpen
  \bibfield  {author} {\bibinfo {author} {\bibfnamefont {B.}~\bibnamefont
  {Hein}}\ and\ \bibinfo {author} {\bibfnamefont {G.}~\bibnamefont {Tanner}},\
  }\bibfield  {title} {\bibinfo {title} {Wave communication across regular
  lattice},\ }\href@noop {} {\bibfield  {journal} {\bibinfo  {journal} {Phys.
  Rev. Lett.}\ }\textbf {\bibinfo {volume} {103}},\ \bibinfo {pages} {260501}
  (\bibinfo {year} {2009}{\natexlab{b}})}\BibitemShut {NoStop}%
\bibitem [{\citenamefont {Barr}\ \emph {et~al.}(2014)\citenamefont {Barr},
  \citenamefont {Proctor}, \citenamefont {Allen},\ and\ \citenamefont
  {Kendon}}]{barr2014}%
  \BibitemOpen
  \bibfield  {author} {\bibinfo {author} {\bibfnamefont {K.}~\bibnamefont
  {Barr}}, \bibinfo {author} {\bibfnamefont {T.}~\bibnamefont {Proctor}},
  \bibinfo {author} {\bibfnamefont {D.}~\bibnamefont {Allen}},\ and\ \bibinfo
  {author} {\bibfnamefont {V.~M.}\ \bibnamefont {Kendon}},\ }\bibfield  {title}
  {\bibinfo {title} {{Periodicity and perfect state transfer in quantum walks
  on variants of cycles}},\ }\href@noop {} {\bibfield  {journal} {\bibinfo
  {journal} {{Quantum Inf. Comput.}}\ }\textbf {\bibinfo {volume} {{14}}},\
  \bibinfo {pages} {{417}} (\bibinfo {year} {{2014}})}\BibitemShut {NoStop}%
\bibitem [{\citenamefont {Štefaňák}\ and\ \citenamefont
  {Skoupý}(2016)}]{stefanak2016}%
  \BibitemOpen
  \bibfield  {author} {\bibinfo {author} {\bibfnamefont {M.}~\bibnamefont
  {Štefaňák}}\ and\ \bibinfo {author} {\bibfnamefont {S.}~\bibnamefont
  {Skoupý}},\ }\bibfield  {title} {\bibinfo {title} {Perfect state transfer by
  means of discrete-time quantum walk on highly symmetric graphs},\ }\href@noop
  {} {\bibfield  {journal} {\bibinfo  {journal} {Phys. Rev. A}\ }\textbf
  {\bibinfo {volume} {94}},\ \bibinfo {pages} {022301} (\bibinfo {year}
  {2016})}\BibitemShut {NoStop}%
\bibitem [{\citenamefont {Štefaňák}\ and\ \citenamefont
  {Skoupý}(2017)}]{stefanak2017}%
  \BibitemOpen
  \bibfield  {author} {\bibinfo {author} {\bibfnamefont {M.}~\bibnamefont
  {Štefaňák}}\ and\ \bibinfo {author} {\bibfnamefont {S.}~\bibnamefont
  {Skoupý}},\ }\bibfield  {title} {\bibinfo {title} {Perfect state transfer by
  means of discrete-time quantum walk on complete bipartite graphs},\
  }\href@noop {} {\bibfield  {journal} {\bibinfo  {journal} {Quantum Inf.
  Process.}\ }\textbf {\bibinfo {volume} {16}},\ \bibinfo {pages} {72}
  (\bibinfo {year} {2017})}\BibitemShut {NoStop}%
\bibitem [{\citenamefont {Zhan}(2019)}]{zhan2019}%
  \BibitemOpen
  \bibfield  {author} {\bibinfo {author} {\bibfnamefont {H.}~\bibnamefont
  {Zhan}},\ }\bibfield  {title} {\bibinfo {title} {{An infinite family of
  circulant graphs with perfect state transfer in discrete quantum walks}},\
  }\href@noop {} {\bibfield  {journal} {\bibinfo  {journal} {{Quantum Inf.
  Process.}}\ }\textbf {\bibinfo {volume} {{18}}},\ \bibinfo {pages} {{369}}
  (\bibinfo {year} {{2019}})}\BibitemShut {NoStop}%
\bibitem [{\citenamefont {Cao}\ \emph {et~al.}(2019)\citenamefont {Cao},
  \citenamefont {Yang}, \citenamefont {Li}, \citenamefont {Dong}, \citenamefont
  {Zhou},\ and\ \citenamefont {Shi}}]{cao2019}%
  \BibitemOpen
  \bibfield  {author} {\bibinfo {author} {\bibfnamefont {W.~F.}\ \bibnamefont
  {Cao}}, \bibinfo {author} {\bibfnamefont {Y.~G.}\ \bibnamefont {Yang}},
  \bibinfo {author} {\bibfnamefont {D.}~\bibnamefont {Li}}, \bibinfo {author}
  {\bibfnamefont {J.~R.}\ \bibnamefont {Dong}}, \bibinfo {author}
  {\bibfnamefont {Y.~H.}\ \bibnamefont {Zhou}},\ and\ \bibinfo {author}
  {\bibfnamefont {W.~M.}\ \bibnamefont {Shi}},\ }\bibfield  {title} {\bibinfo
  {title} {Quantum state transfer on unsymmetrical graphs via discrete-time
  quantum walk},\ }\href@noop {} {\bibfield  {journal} {\bibinfo  {journal}
  {Mod. Phys. Lett. A}\ }\textbf {\bibinfo {volume} {34}},\ \bibinfo {pages}
  {1950317} (\bibinfo {year} {2019})}\BibitemShut {NoStop}%
\bibitem [{\citenamefont {Skoup\'y}\ and\ \citenamefont {\ifmmode
  \check{S}\else \v{S}\fi{}tefa\ifmmode~\check{n}\else
  \v{n}\fi{}\'ak}(2021)}]{skoupy:2022}%
  \BibitemOpen
  \bibfield  {author} {\bibinfo {author} {\bibfnamefont {S.}~\bibnamefont
  {Skoup\'y}}\ and\ \bibinfo {author} {\bibfnamefont {M.}~\bibnamefont
  {\ifmmode \check{S}\else \v{S}\fi{}tefa\ifmmode~\check{n}\else
  \v{n}\fi{}\'ak}},\ }\bibfield  {title} {\bibinfo {title} {Quantum-walk-based
  state-transfer algorithms on the complete $m$-partite graph},\ }\href@noop {}
  {\bibfield  {journal} {\bibinfo  {journal} {Phys. Rev. A}\ }\textbf {\bibinfo
  {volume} {103}},\ \bibinfo {pages} {042222} (\bibinfo {year}
  {2021})}\BibitemShut {NoStop}%
\bibitem [{\citenamefont {Santos}(2022)}]{Santos_2022}%
  \BibitemOpen
  \bibfield  {author} {\bibinfo {author} {\bibfnamefont {R.~A.~M.}\
  \bibnamefont {Santos}},\ }\bibfield  {title} {\bibinfo {title} {Quantum state
  transfer on the complete bipartite graph},\ }\href@noop {} {\bibfield
  {journal} {\bibinfo  {journal} {J. Phys. A: Math. Theor.}\ }\textbf {\bibinfo
  {volume} {55}},\ \bibinfo {pages} {125301} (\bibinfo {year}
  {2022})}\BibitemShut {NoStop}%
\bibitem [{\citenamefont {Štefaňák}\ and\ \citenamefont
  {Skoupý}(2023)}]{stefanak_hypercube_2023}%
  \BibitemOpen
  \bibfield  {author} {\bibinfo {author} {\bibfnamefont {M.}~\bibnamefont
  {Štefaňák}}\ and\ \bibinfo {author} {\bibfnamefont {S.}~\bibnamefont
  {Skoupý}},\ }\bibfield  {title} {\bibinfo {title} {Quantum walk state
  transfer on a hypercube},\ }\href {https://doi.org/10.1088/1402-4896/acf3a2}
  {\bibfield  {journal} {\bibinfo  {journal} {Phys. Scr.}\ }\textbf {\bibinfo
  {volume} {98}},\ \bibinfo {pages} {104003} (\bibinfo {year}
  {2023})}\BibitemShut {NoStop}%
\bibitem [{\citenamefont {Li}\ \emph {et~al.}(2023)\citenamefont {Li},
  \citenamefont {Huang}, \citenamefont {Zhou},\ and\ \citenamefont
  {Yang}}]{li_high-fidelity_2023}%
  \BibitemOpen
  \bibfield  {author} {\bibinfo {author} {\bibfnamefont {D.}~\bibnamefont
  {Li}}, \bibinfo {author} {\bibfnamefont {J.-N.}\ \bibnamefont {Huang}},
  \bibinfo {author} {\bibfnamefont {Y.-Q.}\ \bibnamefont {Zhou}},\ and\
  \bibinfo {author} {\bibfnamefont {Y.-G.}\ \bibnamefont {Yang}},\ }\bibfield
  {title} {\bibinfo {title} {A high-fidelity quantum state transfer algorithm
  on the complete bipartite graph},\ }\href
  {https://doi.org/10.1007/s11128-023-03977-4} {\bibfield  {journal} {\bibinfo
  {journal} {Quantum Inform. Process.}\ }\textbf {\bibinfo {volume} {22}},\
  \bibinfo {pages} {245} (\bibinfo {year} {2023})}\BibitemShut {NoStop}%
\bibitem [{\citenamefont {Huang}\ \emph {et~al.}(2024)\citenamefont {Huang},
  \citenamefont {Li}, \citenamefont {Li}, \citenamefont {Zhou},\ and\
  \citenamefont {Yang}}]{huang_perfect_2024}%
  \BibitemOpen
  \bibfield  {author} {\bibinfo {author} {\bibfnamefont {J.-N.}\ \bibnamefont
  {Huang}}, \bibinfo {author} {\bibfnamefont {D.}~\bibnamefont {Li}}, \bibinfo
  {author} {\bibfnamefont {P.}~\bibnamefont {Li}}, \bibinfo {author}
  {\bibfnamefont {Y.-Q.}\ \bibnamefont {Zhou}},\ and\ \bibinfo {author}
  {\bibfnamefont {Y.-G.}\ \bibnamefont {Yang}},\ }\bibfield  {title} {\bibinfo
  {title} {Perfect state transfer by means of discrete-time quantum walk on the
  complete bipartite graph},\ }\href {https://doi.org/10.1088/1402-4896/ad137a}
  {\bibfield  {journal} {\bibinfo  {journal} {Phys. Scr.}\ }\textbf {\bibinfo
  {volume} {99}},\ \bibinfo {pages} {015110} (\bibinfo {year}
  {2024})}\BibitemShut {NoStop}%
\bibitem [{\citenamefont {Razzoli}\ \emph {et~al.}(2022)\citenamefont
  {Razzoli}, \citenamefont {Bordone},\ and\ \citenamefont
  {Paris}}]{Razzoli_hub_ctqw_2022}%
  \BibitemOpen
  \bibfield  {author} {\bibinfo {author} {\bibfnamefont {L.}~\bibnamefont
  {Razzoli}}, \bibinfo {author} {\bibfnamefont {P.}~\bibnamefont {Bordone}},\
  and\ \bibinfo {author} {\bibfnamefont {M.~G.~A.}\ \bibnamefont {Paris}},\
  }\bibfield  {title} {\bibinfo {title} {Universality of the fully connected
  vertex in laplacian continuous-time quantum walk problems},\ }\href
  {https://doi.org/10.1088/1751-8121/ac72d5} {\bibfield  {journal} {\bibinfo
  {journal} {J. Phys. A: Math. Theor.}\ }\textbf {\bibinfo {volume} {55}},\
  \bibinfo {pages} {265303} (\bibinfo {year} {2022})}\BibitemShut {NoStop}%
\bibitem [{\citenamefont {Novo}\ \emph {et~al.}(2015)\citenamefont {Novo},
  \citenamefont {Chakraborty}, \citenamefont {Mohseni}, \citenamefont {Neven},\
  and\ \citenamefont {Omar}}]{novo2015}%
  \BibitemOpen
  \bibfield  {author} {\bibinfo {author} {\bibfnamefont {L.}~\bibnamefont
  {Novo}}, \bibinfo {author} {\bibfnamefont {S.}~\bibnamefont {Chakraborty}},
  \bibinfo {author} {\bibfnamefont {M.}~\bibnamefont {Mohseni}}, \bibinfo
  {author} {\bibfnamefont {H.}~\bibnamefont {Neven}},\ and\ \bibinfo {author}
  {\bibfnamefont {Y.}~\bibnamefont {Omar}},\ }\bibfield  {title} {\bibinfo
  {title} {Systematic dimensionality reduction for quantum walks: Optimal
  spatial search and transport on non-regular graphs},\ }\href@noop {}
  {\bibfield  {journal} {\bibinfo  {journal} {Sci. Rep.}\ }\textbf {\bibinfo
  {volume} {5}},\ \bibinfo {pages} {13304} (\bibinfo {year}
  {2015})}\BibitemShut {NoStop}%
\end{thebibliography}%

\end{document}